\documentclass[twocolumn]{aastex631} 

\usepackage{gensymb}
\def\Msun{\mbox{${\rm M}_{\odot}$}}
\def\Lsun{\mbox{${\rm L}_{\odot}$}}
\def\Rsun{\mbox{${\rm R}_{\odot}$}}
\usepackage[normalem]{ulem}

\shorttitle{$\kappa$~Dra}
\shortauthors{Klement et al.}

\graphicspath{{./}{figures/}}

\begin{document}

\title{Dynamical masses of the primary Be star and the secondary sdB star \\
in the single-lined binary $\kappa$ Dra (B6\,IIIe)}

\author[0000-0002-4313-0169]{Robert Klement}
\affiliation{The CHARA Array of Georgia State University, Mount Wilson Observatory, Mount Wilson, CA 91023, USA}

\author[0000-0003-1637-9679]{Dietrich Baade}
\affiliation{European Organisation for Astronomical Research in the Southern Hemisphere (ESO), \\ Karl-Schwarzschild-Str.\ 2, 85748 Garching bei M\"unchen, Germany}

\author[0000-0003-1013-5243]{Thomas Rivinius}
\affiliation{European Organisation for Astronomical Research in the Southern Hemisphere (ESO), Casilla 19001, Santiago 19, Chile}

\author[0000-0001-8537-3583]{Douglas R. Gies}
\affiliation{Center for High Angular Resolution Astronomy, Department of Physics and Astronomy,\\ Georgia State University, P.O. Box 5060, Atlanta, GA 30302-5060, USA}

\author[0000-0003-4511-6800]{Luqian Wang}
\affiliation{Center for High Angular Resolution Astronomy, Department of Physics and Astronomy,\\ Georgia State University, P.O. Box 5060, Atlanta, GA 30302-5060, USA}
\affiliation{Yunnan Observatories, CAS, P.O. Box 110, Kunming 650011, Yunnan, China}

\author[0000-0002-2919-6786]{Jonathan Labadie-Bartz}
\affiliation{Homer L. Dodge Department of Physics and Astronomy, University of Oklahoma, 440 W. Brooks Street, Norman, OK 73019, USA}

\author[0000-0002-4808-7796]{Pedro Ticiani dos Santos}
\affiliation{Instituto de Astronomia, Geofísica e Ciências Atmosféricas, Universidade de São Paulo, Rua do Matão 1226, Cidade Universitária, 05508-900 São Paulo, SP, Brazil}

\author[0000-0002-3380-3307]{John D. Monnier}
\affiliation{Department of Astronomy, University of Michigan, 1085 S. University Ave, Ann Arbor, MI 48109, USA}

\author[0000-0002-9369-574X]{Alex C. Carciofi}
\affiliation{Instituto de Astronomia, Geofísica e Ciências Atmosféricas, Universidade de São Paulo, Rua do Matão 1226, Cidade Universitária, 05508-900 São Paulo, SP, Brazil}

\author[0000-0003-2125-0183]{Antoine Mérand}
\affiliation{European Organisation for Astronomical Research in the Southern Hemisphere (ESO), \\ Karl-Schwarzschild-Str.\ 2, 85748 Garching bei M\"unchen, Germany}

\author[0000-0002-2208-6541]{Narsireddy Anugu}
\affiliation{The CHARA Array of Georgia State University, Mount Wilson Observatory, Mount Wilson, CA 91023, USA}

\author[0000-0001-5415-9189]{Gail H. Schaefer}
\affiliation{The CHARA Array of Georgia State University, Mount Wilson Observatory, Mount Wilson, CA 91023, USA}

\author[0000-0002-0493-4674]{Jean-Baptiste Le Bouquin}
\affiliation{Université Grenoble Alpes, CNRS, IPAG, 38000 Grenoble, France}

\author[0000-0001-9764-2357]{Claire L. Davies}
\affiliation{Astrophysics Group, Department of Physics \& Astronomy, University of Exeter, Stocker Road, Exeter, EX4 4QL, UK}

\author[0000-0002-1575-4310]{Jacob Ennis}
\affiliation{Department of Astronomy, University of Michigan, 1085 S. University Ave, Ann Arbor, MI 48109, USA}

\author[0000-0002-3003-3183]{Tyler Gardner}
\affiliation{Department of Astronomy, University of Michigan, 1085 S. University Ave, Ann Arbor, MI 48109, USA}

\author[0000-0001-6017-8773]{Stefan Kraus}
\affiliation{Astrophysics Group, School of Physics and Astronomy, University of Exeter, Stocker Road, Exeter, EX4 4QL, UK}

\author[0000-0001-5980-0246]{Benjamin R. Setterholm}
\affiliation{Department of Astronomy, University of Michigan, 1085 S. University Ave, Ann Arbor, MI 48109, USA}

\author[0000-0001-8837-7045]{Aaron Labdon}
\affiliation{European Organisation for Astronomical Research in the Southern Hemisphere (ESO), Casilla 19001, Santiago 19, Chile}

\begin{abstract}

Because many classical Be stars may owe their nature to mass and angular-momentum transfer in a close binary, the present masses, temperatures, and radii of their components are of high interest for comparison to stellar evolution models.  $\kappa$~Dra is a 61.5-day single-lined binary with a B6\,IIIe primary. With the CHARA Array instruments MIRC/MIRC-X and MYSTIC, we detected the secondary at (approximately photospheric) flux ratios of $1.49\pm0.10$\% and $1.63\pm0.09$\% in the $H$ and $K$ band, respectively.  From a large and diverse optical spectroscopic database only the radial velocity curve of the Be star could be extracted.  However, employing the parallaxes from Hipparcos and Gaia, which agree within their nominal 1-$\sigma$ errors, we could derive the total mass and found component masses of $3.65\pm0.48$~M$_{\odot}$ and $0.426\pm0.043$~M$_{\odot}$ for the Be star and the companion, respectively.  Previous cross-correlation of the observed far-UV spectrum with sdO spectral model templates had not detected a companion belonging to the hot O-type subdwarf (sdO) population known from $\sim20$ earlier-type Be stars. Guided by our full 3D orbital solution, we found a strong cross-correlation signal for a stripped subdwarf B-type companion (far-UV flux ratio of $2.3\pm0.5$\%), enabling the first firm characterization of such a star, and making $\kappa$~Dra the first mid- to late-type Be star with a directly-observed subdwarf companion. 
\end{abstract}


\section{Introduction} \label{sec:intro}

Be stars are extremely rapidly rotating and non-radially pulsating B-type stars of luminosity classes V, IV, and III that possess self-ejected, gaseous disks with  $\sim$Keplerian rotation \citep{2013A&ARv..21...69R}. They constitute about 15 to 20\% of all B-type stars in the local Galactic environment \citep{2013A&ARv..21...69R}. From a stellar evolution perspective, a significant fraction of Be stars are thought to be products of binary interaction and mass and angular-momentum exchange between two initial B-type stars, one of which expands while it evolves beyond the main sequence \citep[MS, e.g.][]{1991A&A...241..419P, 2014ApJ...796...37S, 2021ApJ...908...67S}. The mass transfer from the originally more massive component rejuvenates and spins up the mass gainer which then becomes the classical Be star.  The mass donor loses its envelope (and with it a large fraction of its initial mass), while the hot stripped core remains and is often referred to as an OB-type subdwarf star (sdOB). Further evolution of the system may result in a white dwarf (WD), a neutron star (NS), or a black hole (BH).

The hypothesis that a given Be star is a binary interaction product can be positively supported through (i) the detection of a stripped companion or (ii) the identification of single Be stars as binary merger products or (iii) runaway stars from a former binary disrupted in a supernova explosion.  On the other hand, it can be ruled out by detecting a close stellar companion that is still on the main sequence; this would prove that Be stars can acquire their properties as a consequence of the initial formation conditions \citep{2007A&A...462..683M, 2020A&A...633A.165H} and/or internal structural changes during their main sequence lifetimes \citep{2008A&A...478..467E, 2013A&A...553A..25G}.  Finding a truly single Be star that is not a binary interaction product would lead to the same conclusion.  However, the presence of very faint and low-mass companions such as WDs or sdOBs is difficult to rule out with current observational means \citep[cf.\ the case of the Bn star Regulus with a pre-WD companion,][]{2020ApJ...902...25G}.  Mass transfer products such as Be stars could also be identifiable as blue stragglers, i.e., stars likely rejuvenated by mass transfer, in stellar clusters \citep{2021A&A...652A..70B}.  Furthermore, since most Be stars are nonradial pulsators, long-term monitoring by space photometry can place tight limits on orbital Doppler shifts of the pulsation frequencies \citep[cf.\ ][and references therein]{2018A&A...620A.145B}.  The method is most promising for early-type Be stars, in which p-mode frequencies are found that are relatively isolated and both higher and more stable than the others \citep{2022AJ....163..226L} so that they require shorter observing timespans.  For current facilities, these timespans are too long, but they may be within reach of PLATO \citep{2014ExA....38..249R}.
So far, no model has claimed exclusiveness in its explanation of Be stars, although binary products have been positively identified in many cases.

Currently about 20 Be stars are known to have stripped companions, most of which were confirmed with far UV (FUV) spectroscopy, as it is in the far-UV where the flux ratios are most favorable \citep[][and references therein]{2021AJ....161..248W, 2022ApJ...926..213K}. All of the spectroscopic FUV detections were found to be compatible with sdO nature of the companions, while no firm case of a cooler sdB companion has been presented as of yet. There are also more than 160 confirmed and candidate Be X-ray binaries \citep[BeXRBs,][]{2005A&AT...24..151R}\footnote{List updated at \url{http://xray.sai.msu.ru/~raguzova/BeXcat/}}, which are mostly Be+NS systems that probably evolved in a similar fashion as the Be+sdO binaries but from more massive progenitor systems \citep{2011Ap&SS.332....1R}. The Be primaries occupy a narrow range between O9 and B2 in spectral type \citep{2017A&A...598A..16R}. These systems are conspicuous due to the X-ray emission resulting from the (episodic) accretion of Be disk material onto the compact object, so that the sample is drawn from a much larger (partly extragalactic) volume. White dwarf companions to Be stars proved unexpectedly elusive \citep{1992A&A...265L..41M, 1997ApJ...487..867C}, but several super-soft X-ray emission sources consistent with (early-)Be+WD systems undergoing a Type-II BeXRB outburst were recently detected in the Magellanic Clouds \citep{2020MNRAS.497L..50C, 2021MNRAS.508..781K}. One Be star (MWC~656) was reported to have a BH companion, but this was recently shown to be questionable on the basis of new higher-quality spectra, which rather point towards another Be+sdO system \citep{2022arXiv220812315R}. HD\,93521 is the first candidate post-merger Be star \citep{2022AJ....163..100G}, and members of a Be star runaway population were found using Hipparcos and Gaia astrometric catalogs \citep{2001ApJ...555..364B, 2018MNRAS.477.5261B, 2022ApJS..260...35W}. On the other hand, Be stars have been found missing among B stars at high Galactic latitude \citep{2004AJ....128.2474M, 2006AJ....131.3047M}. Meanwhile, not a single early-type Be star has been confirmed to have a close main sequence companion \citep{2000ASPC..214..668G, 2020A&A...641A..42B}. However, the recently studied case of the B6Ve star $\alpha$~Eri, which is a highly eccentric binary with an early A-type dwarf companion on a seven-year orbit, appears to be the first confirmed case of a Be star that does not owe its nature to mass transfer in a close binary, as the presence of a close stripped companion was ruled out \citep{2022arXiv220907537K}. This implies that two evolutionary channels - single and binary - indeed exist for the formation of Be stars. 

Overall, while many Be stars are known as single-lined spectroscopic (SB1) binaries, very few have a double-lined spectroscopic (SB2) solution or an astrometric orbit, so that estimates of Be star masses are highly uncertain. This deficiency hampers the empirical calibration of binary evolution models. SB2 solutions are rare because the suspected companions are very hard to detect due to their faintness outside the FUV, and the very small angular separations can only be resolved with long-baseline optical/near-IR interferometry. A full SB2 + astrometric solution exists only for two Be stars.  One is the first identified Be+sdO system $\phi$~Per \citep[B1.5Ve,][]{2015A&A...577A..51M}, and the other a special case of the outer star in the hierarchical triple system $\nu$~Gem \citep[B6IIIe,][]{2021ApJ...916...24K}. Preliminary dynamical masses were derived for two additional Be+sdO binaries, but with only SB1 solutions available, these relied on an independent measurement of the parallax \citep{2021ApJ...916...24K}.

$\kappa$~Draconis (= 5~Dra = HR~4787 = HIP~61281 = HD~109387) is a classical Be star and a single-lined spectroscopic binary. The spectral type is B5\,IIIe according to \citet{1982ApJS...50...55S}, B6\,IIIpe according to the Bright Star Catalogue \citep{1991bsc..book.....H, 1995yCat.5050....0H}, and B6 from the temperature derived by \citet{2004A&A...419..607S}. Several SB1 orbital solutions have been published in the last decades, for instance by \citet{1991BAICz..42...39J} and the latest one by \citet{2021RMxAA..57...91S}, all resulting in a circular orbit with an orbital period of 61.55 days. Spectral lines of the secondary were not identified in optical spectra, neither was a signature of a hot sdO companion found in the FUV from IUE spectra \citep{2017ApJ...843...60W}. Using a model spectrum for an effective temperature of 45\,kK, \citeauthor{2017ApJ...843...60W} derived a 5$\sigma$ upper limit on the FUV flux ratio $f_\mathrm{sdO}/f_\mathrm{Be}$ of 0.010. A turndown in the radio spectral slope was detected in $\kappa$~Dra, providing evidence for a truncation of the circumstellar disk due to the orbiting companion \citep{2019ApJ...885..147K}. Previously published interferometric observations by NPOI did not result in a detection of the companion \citep{2021ApJS..257...69H}, nor is there evidence for a more widely separated companion from speckle observations \citep{2020AJ....159..233H}.

Interestingly, hot sdO companions have only been found around early-type Be stars, although the detection should be easier for mid- or late-type Be stars \citep{2019IAUS..346..105R}. It is, therefore, possible that later-type Be stars like $\kappa$~Dra have stripped companions with temperatures comparable to that of the Be primaries, i.e., sdB stars.  Such companions could escape searches optimized for hot sdO companions, as the FUV spectrum morphology differs between sdO and sdB stars. The companion would also remain undetected if it is a late-type MS star and, hence, not the outcome of mass transfer in a close binary, as it would contribute very little flux in the FUV as well as in the optical spectra.

$\kappa$\,Dra was reported to be a source of X-rays based on \textit{Uhuru} observations \citep{1982IAUS...98..353P}, although not at the levels expected for BeXRBs. The detection was not confirmed in more recent surveys with \textit{Rosat} \citep{1996A&AS..118..481B} and \textit{XMM-Newton} \citep{2020MNRAS.493.2511N}, so that there is no indication of a NS companion for $\kappa$~Dra. However, in terms of the X-ray properties, a WD companion remains a possibility, as the above-mentioned facilities were not sensitive enough to detect the expected super-soft X-ray emission \citep[][]{2020MNRAS.497L..50C}. 

$\kappa$~Dra is also known for the rapid variability of its photospheric absorption line profiles, which include travelling subfeatures in the observed line profiles, whose origin is uncertain \citep{1991A&A...246..146H, 2021RMxAA..57...91S}. Not connected to binarity, non-radial pulsations (NRPs) are found in virtually all well-observed Be stars \citep{2003A&A...411..229R, 2022AJ....163..226L} and are made spectroscopically visible by the perturbation of the velocity field across the rotationally-broadened line profiles. In $\kappa$~Dra, the main variability with a reported period of 0.545\,d concerns line width and symmetry and was attributed to a low-order NRP mode \citep{1991A&A...246..146H}. 

As in many other Be stars, longer-term variations in $\kappa$~Dra concern the ratio in strength of the violet ($V$) and red ($R$) peaks of prominent emission lines. These $V/R$ variations are phase-locked to the orbital period \citep{2005Ap&SS.296..173S} and, as in other Be binaries, probably originate from a two-armed ($m=2$) density wave in the disk excited by the companion \citep{2018MNRAS.473.3039P}.  

$\kappa$~Dra also exhibits long-term variations in the strength of its emission lines which have been reported to grow and dissipate with a cycle length of around 22 years \citep{1994A&AS..107..403J, 2004A&A...419..607S}.  In earlier-type Be stars \citep [but see also the case of the B7-8\,IIIe star $\nu$~Pup,][]{2018A&A...620A.145B}, such cyclic behavior - but on shorter timescales - has been seen linked to the interplay of multiple NRP modes \citep{2018pas8.conf...69B}.  Currently, $\kappa$~Dra is in an almost diskless phase, and, according to the mentioned cyclic behavior, the emission is expected to regain strength in the near future. 

In this new study of $\kappa$~Dra, we combine new radial velocities from high-resolution spectroscopy with new astrometric measurements from optical long-baseline interferometry (Sect.~\ref{sec:observations}).  The goals are the determination of the 3-dimensional binary orbit of $\kappa$~Dra (Sect.~\ref{sec:orbit}), a discussion of the fundamental parameters of the primary Be star, and a characterization of the nature of the companion (Sect.~\ref{sec:nature_of_comp}). After a concise summary, Sect.~\ref{sec:conclusions} develops the conclusions as well as goals for future observations.

\section{Observations}
\label{sec:observations}

\subsection{Hipparcos and Gaia astrometry}
\label{sec:observations_hip_gaia}

The parallax measurements for $\kappa$~Dra from the Hipparcos \citep{1997ESASP1200.....E, 2007A&A...474..653V} and Gaia \citep{2016A&A...595A...1G, 2018A&A...616A...1G, 2021A&A...649A...1G} space missions are in good mutual agreement. The re-reduction of the Hipparcos astrometry by \citet{2007A&A...474..653V} should supersede the original reduction, which also considered a single-star solution for $\kappa$~Dra, although the resulting parallaxes agree well withing the formal errors. The distances derived from the  Hipparcos re-reduction, Gaia Data Release 2 (DR2), and Gaia Data Release 3 (DR3) catalogs amount to $150.4^{+8.1}_{-7.3}$\,pc, $140.1^{+6.8}_{-6.2}$\,pc, and $142.7^{+6.3}_{-5.8}$\,pc, respectively (formal parallax errors are the only uncertainties considered here).

Further inspection of the Gaia DR3 data reveals that, with a value of 2.441, the quality flag `renormalised unit weight error' (RUWE) exceeds the threshold of 1.4, above which astrometric solutions are considered degraded \citep{2021A&A...649A...2L}. Therefore, it appears that the sub-mas orbital motions of the primary Be star in $\kappa$~Dra (reported in Sect.~\ref{sec:cp_fitting}), are appreciably affecting the Gaia astrometric solution.  Alternatively, the astrometric solution could be degraded due to detector saturation ($\kappa$~Dra is very bright at $G=3.9$) or variability in the circumstellar environment. In any case, it was found that for RUWE$<3$ the astrometric solution can still be useful within the uncertainties \citep{2021A&A...649A..13M}, and since the agreement between Hipparcos and the individual Gaia releases is good, we adopt the Gaia DR3 parallax and proper motions for the purposes of this study. 

The proper motion of $\kappa$~Dra according to Gaia DR3 \citep{2021A&A...649A...1G} is in good agreement with that measured by Hipparcos \citep{2007A&A...474..653V}, and there are no indications of a proper motion anomaly indicating orbital motions on long time scales \citep{2019A&A...623A..72K, 2022A&A...657A...7K}. The peculiar tangential velocity, obtained after subtracting the Sun's peculiar motion \citep{2010MNRAS.403.1829S} and differential Galactic rotation \citep{1998A&A...331..949M, 1999A&A...345..321M} is $26.7\pm1.2$\,km\,s$^{-1}$. The peculiar space velocity calculated from the systemic velocity of $\kappa$~Dra determined in this work (see Sect.~\ref{sec:orb_solution}) is $33.8\pm1.2$\,km\,s$^{-1}$. These values are in good agreement with those previously determined on the basis of Hipparcos \citep{2001ApJ...555..364B}. Even though the space velocity is close to the threshold of 40\,km\,s$^{-1}$ typically considered for runaway stars \citep{1961BAN....15..265B}, $\kappa$~Dra is a close non-eccentric binary (Sect.~\ref{sec:orb_solution}) with a non-compact companion, and therefore the runaway status bears no significance to the origin of the Be star nature of the primary star.

\subsection{Photometry and spectrophotometry}

\subsubsection{Spectral energy distribution}
\label{sec:obs_sed}

For the basic fitting of the spectral energy distribution (SED), we used the dataset collected by \citet{2019ApJ...885..147K} and complemented it with flux-calibrated UV spectra from the \textit{International Ultraviolet Explorer} \citep[IUE,][]{1978Natur.275..372B}, and near-IR photometry measured by \citet{1991AJ....102.1753D}. A total of 32 IUE spectra were retrieved from the INES database\footnote{\url{http://sdc.cab.inta-csic.es/cgi-ines/IUEdbsMY}}. They were selected to have been taken with the large aperture to ensure reliable flux calibration, but two clear outliers had to be discarded (SWP29665 and LWP08089). In fact, SWP29665 appears to be a mis-identified spectrum of the Be binary 59~Cygni, because it has the same appearance as others of 59~Cygni, and a CCF analysis with a hot 45~kK template reveals the spectral signature of the hot companion at the predicted velocity \citep{2013ApJ...765....2P} for the date of the observation.

The other sources of SED data include the \textit{HPOL} spectropolarimeter\footnote{\url{http://www.sal.wisc.edu/HPOL/}} \citep{2014JAI.....350009D}, the Simbad database\footnote{\url{http://simbad.u-strasbg.fr/simbad/}} \citep{2000A&AS..143....9W}, and the catalogs of the IR space missions \textit{IRAS}\footnote{\url{https://irsa.ipac.caltech.edu/IRASdocs/iras.html}} \citep{1988iras....1.....B, 1994yCat.2125....0J}, \textit{WISE}\footnote{\url{https://www.nasa.gov/mission_pages/WISE/main/index.html}} \citep{2010AJ....140.1868W, 2014yCat.2328....0C}, and \textit{AKARI/IRC}\footnote{\url{https://www.isas.jaxa.jp/en/missions/spacecraft/past/akari.html}} \citep{2010A&A...514A...1I, 2010yCat.2297....0I}.

\subsection{Near-IR interferometry}
\label{sec:obs_interferometry}

Near-IR long-baseline interferometry was secured at the Center for High Angular Resolution Astronomy (CHARA) Array \citep{2005ApJ...628..453T, 2020SPIE11446E..05S} with the beam combiner MIRC \citep{2006SPIE.6268E..1PM} and its upgraded successor MIRC-X \citep{2020AJ....160..158A}.  MIRC(-X) is an image-plane beam combiner operating in the $H$~band at low spectral resolving powers of 50, 100, or 190, resulting in interferometric field of view (FoV) of $\sim50$, $\sim100$, and $\sim200$\,mas, respectively.  In combination with the six CHARA 1-meter telescopes and long baselines of up to $\sim330$\,m, MIRC(-X) achieves angular resolution of down to $\sim0.5$\,milliarcsec (mas) and a very good coverage of the $(u,v)$ plane even with snapshot observations.  Among many other science cases, these properties enable efficient detection of binary companions to bright ($H \lesssim 7.5$) stars with contrast ratios as high as 1:500. The MIRC-X twin instrument MYSTIC \citep{2022SPIE12183E..0BS} has operated at CHARA simultaneously with MIRC-X in the near-infrared $K$~band (with default $R=49$ corresponding to interferometric FoV of $\sim65$\,mas) since 2021 Aug. 

$\kappa$~Dra was observed on six nights in 2012, 2013, and 2021 with different spectral modes (Table~\ref{tab:mircx_log}). On a single night in 2021 (2021~Dec~18), both MIRC-X and MYSTIC data were obtained simultaneously. The star occupies a position in the sky for which 6-telescope delay-line operation is not available at CHARA, and thus all of the datasets were obtained in 5-telescope mode, i.e., with 10 individual baselines (telescopes pairs) and 10 closed triangles (of which only 5 are independent).  Accordingly, each interferometric observation resulted in 10 sets of squared visibilities ({\sc Vis2}), and 10 sets each of closure phases ({\sc CP}) and triple amplitudes ({\sc T3amp}). The data were reduced using the dedicated pipelines for MIRC\footnote{\url{https://www.chara.gsu.edu/tutorials/mirc-data-reduction}} \citep{2007Sci...317..342M} and MIRC-X/MYSTIC\footnote{\url{https://gitlab.chara.gsu.edu/lebouquj/mircx_pipeline.git}} \citep[pipeline version 1.3.5,][]{2020AJ....160..158A}. Some datasets required the removal of edge spectral channels due to highly degraded signal-to-noise ratio. 

Calibrator observations were obtained before and/or after the science target, and were used to correct for atmospheric and instrumental effects in the raw measurements to obtain absolute-calibrated {\sc Vis2}, {\sc CP}, and {\sc T3amp}.  The calibrators were selected using the SearchCal software\footnote{\url{https://www.jmmc.fr/english/tools/proposal-preparation/search-cal/}} \citep{2016A&A...589A.112C} and are listed along with their uniform disk (UD) diameters in Table~\ref{tab:calibrators}. The calibrator diameters were adopted from the JMMC catalog of stellar diameters \citep{2014ASPC..485..223B, 2017yCat.2346....0B}. 

\begin{deluxetable}{lCC}
\tablecaption{Interferometric calibrators}\label{tab:calibrators}
\tablewidth{0pt}
\tablehead{
\colhead{Calibrator} & \colhead{UD diam. ($H$)} & \colhead{UD diam. ($K$)} \\
\nocolhead{Target} & \colhead{[mas]} & \colhead{[mas]} 
}
\startdata
HD~98499 & 0.709\pm0.079 & 0.712\pm0.079\\
HD~104986 & 0.531\pm0.013 & 0.533\pm0.013\\
HD~106925 & 0.306\pm0.007 & 0.308\pm0.007\\
HD~108399 & 0.763\pm0.066 & 0.767\pm0.066\\
HD~112609 & 0.391\pm0.010 & 0.392\pm0.010\\
HD~116285 & 0.482\pm0.012 & 0.484\pm0.012\\
HD~117113 & 0.511\pm0.012 & 0.513\pm0.012\\
HD~118788 & 0.619\pm0.053 & 0.621\pm0.053\\
\enddata
\end{deluxetable}


The absolute calibration of {\sc Vis2} and {\sc T3amp} is more susceptible to atmospheric turbulence than that of {\sc CP}, as the latter is a differential measurement explicitly designed to cancel the turbulence-induced terms in the interferometric phase. Unfortunately, several of our epochs were affected by bad atmospheric conditions; a combination of poor seeing, spikes in humidity and/or wind gusts made for unstable fringes on 2012~Apr~28, 2021~Mar~6, 2021~May~20, and to a lesser extent on 2021~Dec~18. While the {\sc CP} precision turned out to be sufficient even on these nights, {\sc Vis2} and {\sc T3amp} had to be completely discarded for the three most affected ones. For 2021~Dec~18, {\sc Vis2} and {\sc T3amp} had to be discarded from the MIRC-X part of the dataset, while the MYSTIC part could be used in full. The calibrated OIFITS files corresponding to all MIRC(-X) and MYSTIC observations analyzed in this work will be available in the Optical Interferometry Database\footnote{\url{http://oidb.jmmc.fr/index.html}} \citep{2014SPIE.9146E..0OH} and the CHARA Data Archive\footnote{\url{https://www.chara.gsu.edu/observers/database}}.

\begin{deluxetable*}{clCCcccc}
\tablecaption{Log of MIRC(-X) and MYSTIC observations\label{tab:mircx_log}}
\tablewidth{0pt}
\tablehead{
\colhead{HJD $- 2400000.5$} & \colhead{UT date} & \colhead{{\sc Vis2}} & \colhead{{\sc CP}/{\sc T3AMP}} & \colhead{Configuration} & \colhead{Instrument} & \colhead{Spectral mode} 
}
\startdata
56045.2064 & 2012 Apr 28    & \nodata  & 2310 & E1-W2-W1-S2-E2 & MIRC & H-Prism 50\\
56410.2534 & 2013 Apr 28    & 4725  & 4585  & E1-W2-W1-S2-E2 & MIRC & H-Prism 50\\
59279.5134 & 2021 Mar 6    & \nodata & 17024  & E1-W1-S2-S1-E2 & MIRC-X & H-Grism 190\\
59354.2334 & 2021 May 20    & \nodata  & 2840  & E1-W2-W1-S2-E2 & MIRC-X & H-Prism 102\\
59377.1954 & 2021 Jun 12    & 4867  & 4650 & E1-W2-W1-S2-E2 & MIRC-X & H-Grism 190\\
59566.5107 & 2021 Dec 18    & \nodata  & 1865 & E1-W2-W1-S2-E2 & MIRC-X & H-Prism 50\\
59566.5107 & 2021 Dec 18    & 1944  & 1920 & E1-W2-W1-S2-E2 & MYSTIC & K-Prism 49\\
\enddata
\tablecomments{The columns {\sc VIS2} and {\sc CP}/{\sc T3AMP} give the total number of measurements in $30$-second intervals.}
\end{deluxetable*}

\subsection{Optical spectroscopy}

We assembled over 500 individual spectra from the archives of various spectrographs as well as from the database of Be-star spectra mostly taken by amateur astronomers,  BeSS\footnote{\url{http://basebe.obspm.fr/basebe/}} \citep{BeSS}.  Heliocentric correction was applied to the wavelength scale of all spectra, and all dates were converted to heliocentric Julian dates (HJD).  A significant number of the spectra was originally obtained for spectropolarimetric purposes and to study short-term line profile variations. Therefore, these datasets cluster in few nights each and have a high signal-to-noise ratio (SNR). A subset of 15 {\sc Heros} blue and 15 {\sc Heros} red spectra was previously analyzed by \citet{2004A&A...419..607S, 2005Ap&SS.296..173S, 2021RMxAA..57...91S}.  The basic properties of the spectra, including the resolving power, SNR, and spectral coverage, are summarized in Table~\ref{tab:spectroscopy}, and the spectrographs are briefly introduced in the following. 

BeSS. The available echelle spectra include professional spectra from MUSICOS, as well as amateur spectra taken with different instruments. MUSICOS was a fiber-fed spectrograph for multi-site observations \citep{1992A&A...259..711B}. The quality of these spectra varies and their wavelength calibration might be unreliable. 

ELODIE. This was an echelle spectrograph installed at the Observatoire de Haute-Provence 1.93m reflector in south-eastern France \citep{1996A&AS..119..373B}. Two {\sc Elodie} spectra were recovered from the archive\footnote{\url{http://atlas.obs-hp.fr/elodie/}} \citep{2004PASP..116..693M}.

ESPaDOnS / Narval. ESPaDOnS\footnote{\url{http://www.ast.obs-mip.fr/projets/espadons/espadons.html}} \citep{2003ASPC..307...41D, 2006ASPC..358..362D, 2016MNRAS.456....2W} and Narval\footnote{\url{http://www.ast.obs-mip.fr/projets/narval/}} \citep{2016MNRAS.456....2W} are twin instruments with spectro-polarimetric capabilities.  Data from these instruments are the best-quality spectroscopic datasets available for $\kappa$~Dra.  

{\sc Flash} / {\sc Heros} / {\sc Flash2}. The Fiber-Linked Astronomical Spectrograph of Heidelberg or {\sc Flash} \citep{1988IAUS..132....9M} was a portable fiber-linked echelle spectrograph with a single camera. The Heidelberg Extended Range Optical Spectrograph or {\sc Heros}\footnote{\url{https://www.lsw.uni-heidelberg.de/projects/instrumentation/Heros/}} \citep{1998RvMA...11..177K, 2000ASPC..214..356S} was an upgrade of {\sc Flash} with a thinner fiber (and hence higher spectral resolution) and with two spectral channels (blue and red). {\sc Flash2} was a phase of {\sc Flash} with the thinner fiber but when the blue camera was out of commission and the wavelength range was adjusted for observation with the red arm only. Descriptions of these instruments and the reduction procedure can also be found in \citet{2001JAD.....7....5R} and references therein. 

\begin{deluxetable*}{lccCCCC}
\tablecaption{Spectroscopic datasets\label{tab:spectroscopy}}
\tablewidth{0pt}
\tablehead{
\colhead{Instrument/Detector} & \colhead{Telescope} & \colhead{MJD} & \colhead{Number} & \colhead{Resolving power} & \colhead{SNR} & \colhead{Spectral coverage (\AA)}
}
\startdata
BeSS database    & Multiple     & 54904--56494 & 91  & $\sim$11000 & 100-400 & 4300-6900 \\
ELODIE     & OHP 1.9\,m     & 51570--51572 & 2   & 42000 & 300-400 & 3850-6800 \\
ESPaDOnS   & CFHT 3.6\,m    & 53512--55229 & 104 & 68000 & 400-600 & 3700-10050 \\
{\sc Flash}      & Tautenburg 2.0\,m     & 48261--48443 & 195 & 12000 & 100-400 & 4050-6780\\
{\sc Flash2}     & Wendelstein 80\,cm  & 51680--51757 & 21  & 20000 & 100-400 & 4050-6780\\
{\sc Heros} blue & Ondřejov 2.0\,m & 51899--52714 & 45  & 20000 & 100-300 & 3500-5500  \\
{\sc Heros} red  & Ondřejov 2.0\,m & 51899--52714 & 53  & 20000 & 100-300 & 5600-8600  \\
SWP & IUE & 43953--48191 & 24  & 7500 & 20 & 1150-1980\\
LWP/LWR & IUE & 43953--47272 & 6 & 12500 & 20 & 1850-3350 \\
MUSICOS    & TBL 2.0\,m        & 51897--52265 & 26  & 35000 & 300-500 & 3800-8700\\
Narval     & TBL 2.0\,m        & 54105--55949 & 64  & 65000 & 400-600 & 3700-10050\\
\enddata
\end{deluxetable*}

\section{Orbital analysis}
\label{sec:orbit}

\subsection{Measurement of astrometric positions of the companion from interferometry}
\label{sec:cp_fitting}

\begin{deluxetable*}{CCCCCCCCCCc}
\tablecaption{Relative astrometric positions of the companion. \label{tab:detections}}
\tablewidth{0pt}
\tablehead{
\colhead{HJD$- 2400000.5$} & \colhead{$\rho$} & \colhead{PA}  & \colhead{$\Delta$RA} & \colhead{$\Delta$DEC} & \colhead{$\sigma$-$a$} & \colhead{$\sigma$-$b$} & \colhead{$\sigma$-PA} &
\colhead{$f$} & \colhead{UD diam.} & \colhead{Band} \\ 
\nocolhead{HJD $- 2400000.5$} & \colhead{[mas]} & \colhead{[\degree]} & \colhead{[mas]} & \colhead{[mas]}& \colhead{[mas]} & \colhead{[mas]} & \colhead{[\degree]} & \colhead{[\% primary]} & \colhead{[mas]} & \nocolhead{Data}
}
\startdata
\hline
56045.206 & 3.370 & 287.994  & -3.208 &  1.042 & 0.045 & 0.029 & 94.6 & 1.29\pm0.08 & 0.45\tablenotemark{a} &  H \\
56410.253 & 3.406 & 307.22  & -2.712 &  2.060 & 0.054 & 0.045 & 112.6 & 1.22\pm0.13 & 0.45\pm0.02  &  H \\
59279.510 & 3.213 &  97.733  &  3.183 & -0.433 & 0.023 & 0.019 & 158.2 & 1.33\pm0.06 & 0.354\tablenotemark{a} &  H \\
59354.234 & 2.468 &   0.971  &  0.040 &  2.458 & 0.044 & 0.042 & 86.1 & 1.6\pm0.3 & 0.354\tablenotemark{a} &  H \\ 
59377.197 & 2.594 & 247.518 & -2.397 & -0.992 & 0.031 & 0.016 & 127.2 & 1.59\pm0.09 & 0.354\pm0.004 &  H \\ 59566.509 & 2.180 & 210.292 & -1.100 & -1.882 & 0.037 & 0.017 & 205.6 & 2.01\pm0.13 & 0.354\tablenotemark{a} &  H \\ 
59566.509 & 2.228 & 211.400 & -1.161 & -1.902 & 0.048 & 0.019 & 135.9 & 1.49\pm0.10 & 0.405\pm0.016 &  K  \\
\enddata
\tablenotetext{a}{Fixed.}
\tablecomments{$\rho$ is the angular separation between the components, PA is the position angle (from North to East), $\Delta$RA and $\Delta$DEC are the companion coordinates relative to the primary, $\sigma$-$a$ and $\sigma$-$b$ are the major and minor axes of the error ellipse, respectively, $\sigma$-PA is the position angle of the error ellipse (from North to East), $f$ is the secondary-to-primary flux fraction, and UD diam.\ is the diameter of the primary (including any contribution from the disk).
}
\end{deluxetable*}

We fitted geometric models to the interferometric data with the open-source Python code \textit{PMOIRED}\footnote{\url{https://github.com/amerand/PMOIRED}} \citep[Parametric Modeling of Optical InteRferomEtric Data,][]{2022arXiv220711047M}. \textit{PMOIRED} is an analysis tool for interferometry, which relies on semi-analytical expressions for complex visibilities to construct geometrical models made up from building blocks such as uniform disks and Gaussians. For the companion detection, we used the dedicated open-source Python code \textit{ CANDID}\footnote{\url{https://github.com/amerand/CANDID}, \url{https://github.com/agallenne/GUIcandid}} \citep[Companion Analysis and Non-Detection in Interferometric Data,][]{2015A&A...579A..68G}, which includes corrections for the effect of bandwidth smearing on the derived flux ratios. The uncertainties of the fitted parameters were determined using a data resampling (bootstrapping) algorithm and include the instrumental wavelength calibration uncertainty of 0.5\% \citep{2020AJ....160..158A}.

\begin{figure*}
   \centering
   \includegraphics[width=1.0\textwidth]{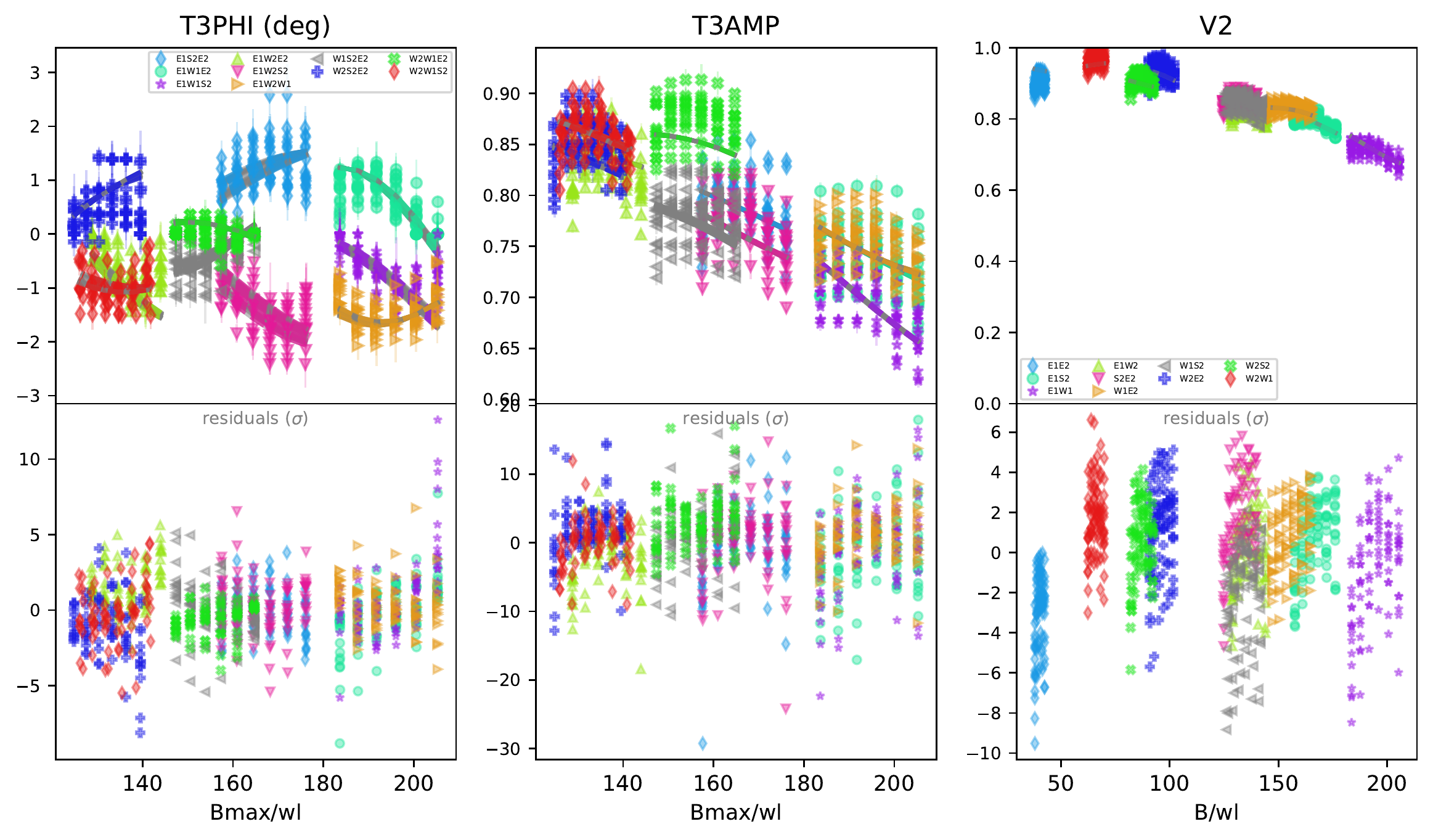}
   \caption{\label{fig:show} \textit{PMOIRED} output plot showing the MIRC-X interferometric data set of $\kappa$~Dra taken on 2021 Jun 12 and the corresponding model fit. The two panels from the left show the {\sc CP} (T3PHI) and {\sc T3AMP} versus the maximum baseline (in units of $10^6\, \mathrm{rad}^{-1}$) for all 10 baseline triangles identified in the legend at the top left, and the right panel shows {\sc Vis2} versus the baseline. The bottom panels show the residuals in units of $\sigma$. The best-fit PMOIRED model consists of a primary star represented by a UD, and a point-source secondary star separated by $\sim2.6$\,mas. The MIRC-X grism data ($R$=190) were median-filtered to decrease $R$ to 50, which limits the (bandwidth-smearing) field of view to $\pm50$\,mas. The plots with the final fits to all seven datasets are available in the online journal.
   }
\end{figure*}

The $\kappa$~Dra binary system was first represented by a UD for the Be primary (including its disk) and a point source for the secondary. A grid search with \textit{CANDID} was used to find the relative position ($\Delta$RA, $\Delta$DEC) and the relative flux $f$ of the secondary (expressed as a percentage of the primary flux) for each epoch. For the epochs with available {\sc Vis2} and {\sc T3Amp}, the primary UD diameter was included as a free parameter, while it was kept fixed for the others. The companion was successfully detected and its position was determined for all seven datasets, and the results are summarized in Table~\ref{tab:detections}. It should be noted that the $\chi^2$ map resulting from the grid search of the MYSTIC dataset - when using only the CP - shows two minima of an equal depth. While one of the minima agrees with the corresponding MIRC-X position and is close to the MIRC-X flux ratio, the other is clearly offset at double the separation and double the flux ratio. Consequently, we adopt the former position as the correct one. The binary fit to the best-quality dataset (MJD\,59377.197) is shown in Fig.~\ref{fig:show}, while the plots with the final fits to all seven datasets are available in the online journal.

\figsetstart
\figsetnum{1}
\figsettitle{Model fits to interferometric data.  
}

\figsetgrpstart
\figsetgrpnum{1.1}
\figsetgrptitle{MIRC 2012~Apr~18}
\figsetplot{PMOIRED_2012Apr28_MIRC_L2.kap_Dra.2012Apr28_tcoh0017ms_GHS_2019Feb28.XCHAN.cal_flag.SPLIT.oifits_SHOW.pdf}
\figsetgrpnote{\textit{PMOIRED} output plot showing the MIRC-X interferometric data set of $\kappa$~Dra taken on 2021~Apr~28 and the corresponding model fit. The two panels from the left show the {\sc CP} (T3PHI) and {\sc T3AMP} versus the maximum baseline (in units of $10^6\, \mathrm{rad}^{-1}$) for all 10 baseline triangles identified in the legend at the top left, and the right panel shows {\sc Vis2} versus the baseline. The bottom panels show the residuals in units of $\sigma$. The best-fit PMOIRED model consists of a primary star represented by a UD, and a point-source secondary star.}
\figsetgrpend

\figsetgrpstart
\figsetgrpnum{1.2}
\figsetgrptitle{MIRC 2013~Apr~18}
\figsetplot{PMOIRED_2013Apr28_MIRC_L2.kap_Dra.2013Apr28_tcoh0075ms_GHS_2019Feb28.XCHAN.cal_flag.SPLIT.oifits_SHOW.pdf}
\figsetgrpnote{\textit{PMOIRED} output plot showing the MIRC-X interferometric data set of $\kappa$~Dra taken on 2013~Apr~28 and the corresponding model fit. The two panels from the left show the {\sc CP} (T3PHI) and {\sc T3AMP} versus the maximum baseline (in units of $10^6\, \mathrm{rad}^{-1}$) for all 10 baseline triangles identified in the legend at the top left, and the right panel shows {\sc Vis2} versus the baseline. The bottom panels show the residuals in units of $\sigma$. The best-fit PMOIRED model consists of a primary star represented by a UD, and a point-source secondary star. }
\figsetgrpend

\figsetgrpstart
\figsetgrpnum{1.3}
\figsetgrptitle{MIRC-X 2021~Mar~6}
\figsetplot{PMOIRED_MIRCX_L2.2021Mar06.kap_Dra.MIRCX_IDL.1.SPLIT_median5.oifits_SHOW.pdf}
\figsetgrpnote{\textit{PMOIRED} output plot showing the MIRC-X interferometric data set of $\kappa$~Dra taken on 2021~Mar~6 and the corresponding model fit. The two panels from the left show the {\sc CP} (T3PHI) and {\sc T3AMP} versus the maximum baseline (in units of $10^6\, \mathrm{rad}^{-1}$) for all 10 baseline triangles identified in the legend at the top left, and the right panel shows {\sc Vis2} versus the baseline. The bottom panels show the residuals in units of $\sigma$. The best-fit PMOIRED model consists of a primary star represented by a UD, and a point-source secondary star. }
\figsetgrpend

\figsetgrpstart
\figsetgrpnum{1.4}
\figsetgrptitle{MIRC-X 2021~May~20}
\figsetplot{PMOIRED_MIRCX_L2.2021May20.kap_Dra.MIRCX_IDL.1.SPLIT.oifits_SHOW.pdf}
\figsetgrpnote{\textit{PMOIRED} output plot showing the MIRC-X interferometric data set of $\kappa$~Dra taken on 2021~May~20 and the corresponding model fit. The two panels from the left show the {\sc CP} (T3PHI) and {\sc T3AMP} versus the maximum baseline (in units of $10^6\, \mathrm{rad}^{-1}$) for all 10 baseline triangles identified in the legend at the top left, and the right panel shows {\sc Vis2} versus the baseline. The bottom panels show the residuals in units of $\sigma$. The best-fit PMOIRED model consists of a primary star represented by a UD, and a point-source secondary star. }
\figsetgrpend

\figsetgrpstart
\figsetgrpnum{1.5}
\figsetgrptitle{MIRC-X 2021~Jun~12}
\figsetplot{PMOIRED_MIRCX_L2.2021Jun12.kap_Dra.MIRCX_IDL.1.SPLIT_median5.oifits_SHOW.pdf}
\figsetgrpnote{\textit{PMOIRED} output plot showing the MIRC-X interferometric data set of $\kappa$~Dra taken on 2021~Jun~12 and the corresponding model fit. The two panels from the left show the {\sc CP} (T3PHI) and {\sc T3AMP} versus the maximum baseline (in units of $10^6\, \mathrm{rad}^{-1}$) for all 10 baseline triangles identified in the legend at the top left, and the right panel shows {\sc Vis2} versus the baseline. The bottom panels show the residuals in units of $\sigma$. The best-fit PMOIRED model consists of a primary star represented by a UD, and a point-source secondary star. }
\figsetgrpend

\figsetgrpstart
\figsetgrpnum{1.6}
\figsetgrptitle{MIRC-X 2021~Dec~18}
\figsetplot{PMOIRED_MIRCX_L2.2021Dec18.kap_Dra.MIRCX_IDL.1.SPLIT.oifits_SHOW.pdf}
\figsetgrpnote{\textit{PMOIRED} output plot showing the MIRC-X interferometric data set of $\kappa$~Dra taken on 2021~Dec~18 and the corresponding model fit. The two panels from the left show the {\sc CP} (T3PHI) and {\sc T3AMP} versus the maximum baseline (in units of $10^6\, \mathrm{rad}^{-1}$) for all 10 baseline triangles identified in the legend at the top left, and the right panel shows {\sc Vis2} versus the baseline. The bottom panels show the residuals in units of $\sigma$. The best-fit PMOIRED model consists of a primary star represented by a UD, and a point-source secondary star. }
\figsetgrpend

\figsetgrpstart
\figsetgrpnum{1.7}
\figsetgrptitle{MYSTIC 2021~Dec~18}
\figsetplot{PMOIRED_MYSTIC_L2.2021Dec18.kap_Dra.MIRCX_IDL.1.SPLIT.oifits_SHOW.pdf}
\figsetgrpnote{\textit{PMOIRED} output plot showing the MYSTIC interferometric data set of $\kappa$~Dra taken on 2021~Dec~18 and the corresponding model fit. The two panels from the left show the {\sc CP} (T3PHI) and {\sc T3AMP} versus the maximum baseline (in units of $10^6\, \mathrm{rad}^{-1}$) for all 10 baseline triangles identified in the legend at the top left, and the right panel shows {\sc Vis2} versus the baseline. The bottom panels show the residuals in units of $\sigma$. The best-fit PMOIRED model consists of a primary star represented by a UD, and a point-source secondary star. }
\figsetgrpend

\figsetend

\begin{figure}
   \centering
   \includegraphics[width=0.5\textwidth]{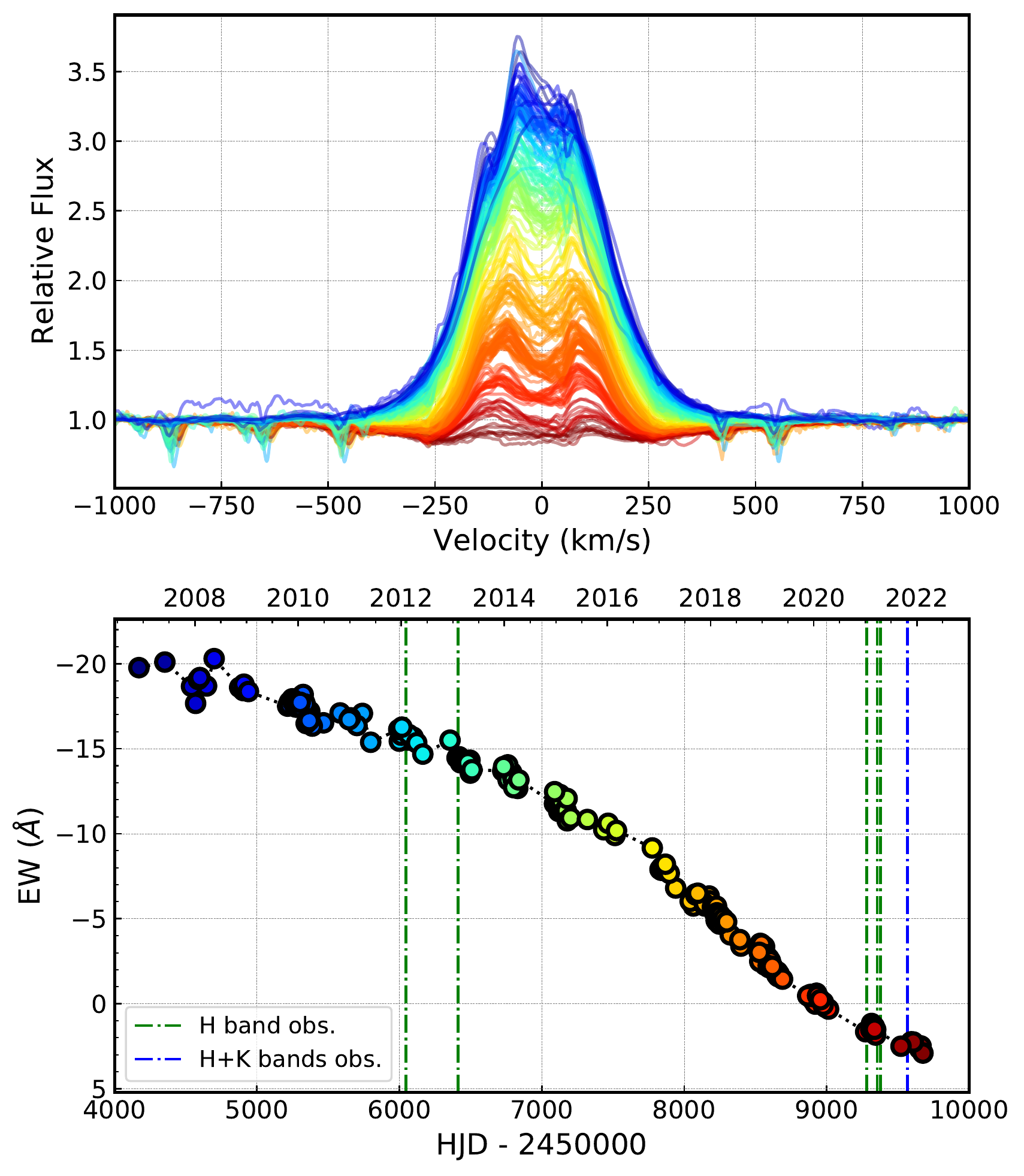}
   \caption{\label{fig:h-alpha} \textit{Upper:} H$\alpha$ line profiles recovered from the BeSS spectral database. The colors correspond to dates as shown in the lower panel. \textit{Lower:} Equivalent width (EW) measured for all H$\alpha$ profiles. A continuous decrease in emission strength is apparent since around 2010 and is still ongoing. Vertical lines show the epochs of our interferometric observations.
   }
\end{figure}

The resulting UD diameter values in the $H$~band show an apparent decrease in size between the years 2013 and 2021, and a larger size in the $K$~band relative to the $H$~band in 2021. The $H$-band flux ratios indicate an increase from 2012/2013 to 2021 when comparing the corresponding averages ($1.25\pm0.08$\% and $1.63\pm0.09$\%), as well as only the best epochs with complete datasets (MJD\,56410.253 and MJD\,59377.197). Finally, the flux ratio in the $K$~band is slightly lower than the average $H$~band flux ratio in 2021, although they agree within the error bars. The apparent decrease in size coupled with a small increase in the relative flux of the companion most likely reflects the significant disk dissipation that had occurred between 2013 and 2021, as evidenced by the decreasing line emission in the spectra (Fig.~\ref{fig:h-alpha}). The larger size in the $K$~band probably reflects what is expected for Be star disks due to the increasing free-free and bound-free opacity in ionized, gaseous environments \citep{2015MNRAS.454.2107V}, i.e., the disk appearing larger at longer wavelengths, although the line emission from the dissipating disk was already very low at this time. The lower flux ratio in the $K$~band relative to the $H$~band points towards a companion that is hotter than the primary Be star, although the higher excess emission from the remaining disk could also contribute to the same effect. 

With the companion positions determined for each epoch, we used \textit{PMOIRED} to include an additional model component - inclined Gaussian - to represent the circumstellar disk of the primary Be star, and attempted to constrain its parameters by fitting the three full datasets. The UD component now representing the Be star photosphere was fixed at the near-photospheric measured diameter of 0.354\,mas. Only the recent better quality MIRC-X and MYSTIC data resulted in a constrained inclination and PA of the Gaussian disk component (Table~\ref{tab:disk_params}), although the overall fit to the data is not significantly improved. This suggests that these parameters are not constrained well by the data, most likely due to the fact that the disk has almost dissipated by 2021.

\begin{deluxetable}{ChCCc}
\tablecaption{Parameters of the disk around the Be star}\label{tab:disk_params}
\tablewidth{0pt}
\tablehead{
\colhead{HJD $- 2400000.5$} & \nocolhead{FWHM} & \colhead{$i$} & \colhead{PA} & \colhead{Band} \\
\nocolhead{Target} & \nocolhead{[mas]} & \colhead{[\degree]} & \colhead{[\degree]}
}
\startdata
59377.197 & 0.223\pm0.002 & 40.5\pm5.1 & 133.0\pm5.8 & H \\
59566.509 & 0.246\pm0.003 & 34.2\pm9.8 & 119\pm18 & K \\
\enddata
\end{deluxetable}

\subsection{Radial-velocity measurements in spectra}
\label{sec:rv_measurement}

The spectrum of $\kappa$~Dra is dominated by rotationally broadened absorption lines from the primary B6\,IIIe component.  It is overlaid with generally double-peaked emission lines originating from the circumstellar disk; their separations are consistent with an intermediate to low inclination.  No lines attributable to the secondary component are visible. 
The rotational broadening and variability of the line profiles due to NRPs hinder the accuracy of radial-velocity (RV) measurements of the primary component that would be achievable for stars with narrow and invariant lines. Since the disk of $\kappa$~Dra is not viewed close to edge-on and the line of sight does not intersect the disk, there are no narrow shell absorption lines that could be used as a proxy of the stellar RV. Nevertheless, using a large amount of spectra and measuring the RVs separately from several different lines should alleviate possible systematic errors in the RV measurements, as the effects of the rapid variability should average out and only lead to possibly increased scatter.

Therefore, for the purpose of precise RV determinations of the primary Be star, we searched for prominent absorption lines uncontaminated by line emission from the disk and without other obvious circumstances affecting the line shape. Five lines were finally selected following careful inspection: \ion{He}{1}\,$\lambda4026$, \ion{He}{1}\,$\lambda4471$, \ion{Mg}{2}\,$\lambda4481$, \ion{He}{1}\,$\lambda4713$, and \ion{He}{1}\,$\lambda6678$. Even though \ion{Mg}{2}\,$\lambda4481$ and \ion{He}{1}\,$\lambda6678$ are in hotter Be stars often contaminated by emission from the disk, in the case of $\kappa$~Dra there appears to be no emission component in these lines. After normalization of the surrounding continuum, the line profiles were fitted with a Gaussian profile to determine the RV shift at each epoch.  Each measurement was visually checked, and poor fits resulting in clear outliers were discarded.  With the exception of \ion{He}{1}\,$\lambda4713$, all of the selected lines are affected by blends with forbidden transitions of unknown strength, so that the systemic velocity can be most accurately determined from this line \citep{1982bsww.book.....U}.

The spectroscopic datasets used in this study are of heterogeneous quality. The 168 spectra from ESPaDOnS and Narval have the highest quality, resolution, and SNR and yield the most precise RV measurements.  The ELODIE dataset is the second best in terms of spectra quality, but only two spectra are available.  The 25 observations with MUSICOS and the 91 amateur spectra from the BeSS database have comparable resolution to those mentioned above; however, the spectral calibration may be less reliable, as MUSICOS was not optimized for this purpose, and the quality control of the BeSS data reduction procedures may be less stringent. The advantage of the BeSS spectra is that they are more uniformly distributed with orbital phase than the other datasets. The {\sc Flash/Heros/Flash2} spectra are only partly usable for RV measurements due to a much lower resolution. Following an exploration of the influence of discarding lower-quality spectra on the resulting scatter, we conservatively discarded all RV measurements from spectral segments with SNR$<200$ per resolution element. This resulted in a significant reduction of useful spectra for the lower quality datasets from BeSS and {\sc Flash/Heros/Flash2}. 

For each instrument of origin, the scatter of individual RV measurements at similar epochs was used to estimate their 1-$\sigma$ uncertainties. As expected from the data quality, the scatter is the lowest for ESPaDOnS and Narval, and increases by a factor of two to three for the other data sources. The following values were finally adopted: 3\,km\,s$^{-1}$ for ESPaDOnS/Narval, 6\,km\,s$^{-1}$ for ELODIE, and 9\,km\,s$^{-1}$ for the remaining instruments.

\subsection{Orbital Solution}
\label{sec:orb_solution}

The orbital solution was obtained using the code {\sc Fotel} \citep{2004PAICz..92....1H}, which can combine various datasets, such as photometry (for eclipsing binaries), RVs (for spectroscopic binaries), and astrometric orbital positions (for astrometric/interferometric binaries). In the case of $\kappa$~Dra, we combined the RVs measured in the spectra with the astrometric positions measured from near-IR interferometry. For the weights of individual measurements, required for the orbital solution by {\sc Fotel} at input, the inverse of the corresponding measurement uncertainties were adopted. To ensure equal influence of the astrometry and the RVs, the relative weight of the two measurement sets was adjusted so that they contributed approximately equal amounts to the total residuals.  

In addition to the stellar coordinates and proper motion (Sect.~\ref{sec:observations_hip_gaia}), a full dynamical description of a binary system requires the knowledge of ten parameters.  Four of these - the period ($P$), the epoch of periastron passage ($T_0$), the eccentricity ($e$), and the longitude of periastron ($\omega$) - can be determined independently from both the astrometry and the RVs.  On the other hand, only the RVs are sensitive to the line-of-sight motions, i.e., the velocity semi-amplitudes of the two components ($K_1$, $K_2$) and the systemic velocity ($\gamma$), while astrometry is the only source of information about the orbital inclination ($i$), the position angle of the line of nodes ($\Omega$, although RVs are needed to distinguish the ascending and the descending node), and the angular semi-major axis of the orbit \citep[$a''$, e.g.,][]{1973bmss.book.....B}.

With measurements of all ten parameters, the absolute size of the orbit ($a$) can be calculated from the $K$-semi-amplitudes, $i$, and $P$, the distance to the system ($D$) from a comparison of the absolute and angular size of the orbit, and finally the mass of the system from the third Kepler law.  Masses of the individual components ($M$) are then found from the ratio of the $K$-semi-amplitudes.  This ideal case corresponds to an astrometric binary that is also a double-lined spectroscopic binary (SB2), where it is possible to obtain the $K$ semi-amplitudes of both components from composite spectra.  For $\kappa$~Dra, however, which is a single-lined spectroscopic binary (SB1), an independent measurement of the distance $D$ is not possible, and for the determination of the component masses we have to rely on the parallax measurement (Sect.\,\ref{sec:observations_hip_gaia}).

In the first solution, we only used the best-precision RVs from ESPaDOnS and Narval -- a total of 839 from all five spectral lines (Sect.\,\ref{sec:rv_measurement}) -- alongside the astrometry. Fixing $e=0$ to enforce a circular orbit favored by earlier studies (Sect.\,\ref{sec:intro};  Table~\ref{tab:orb_pars}) gave a good fit to both data sets and a well-constrained set of orbital parameters.  Adding $e$ and $\omega$ as free parameters did not result in an elliptic orbit, so that the data are compatible with an infinitesimally low eccentricity and undefined $\omega$. The binary parameters obtained are listed in the first column of Table~\ref{tab:orb_pars}, where they are also compared to the results from older spectroscopic studies, showing very good agreement in the period ($P$), and good agreement in the RV semi-amplitude of the Be star ($K_\mathrm{Be}$). The rms of the $(O-C)$ residuals of the ESPaDOnS and Narval RV measurements is $2.6$\,km\,s$^{-1}$. 

\begin{figure*}
   \centering
   \includegraphics[width=1.0\textwidth]{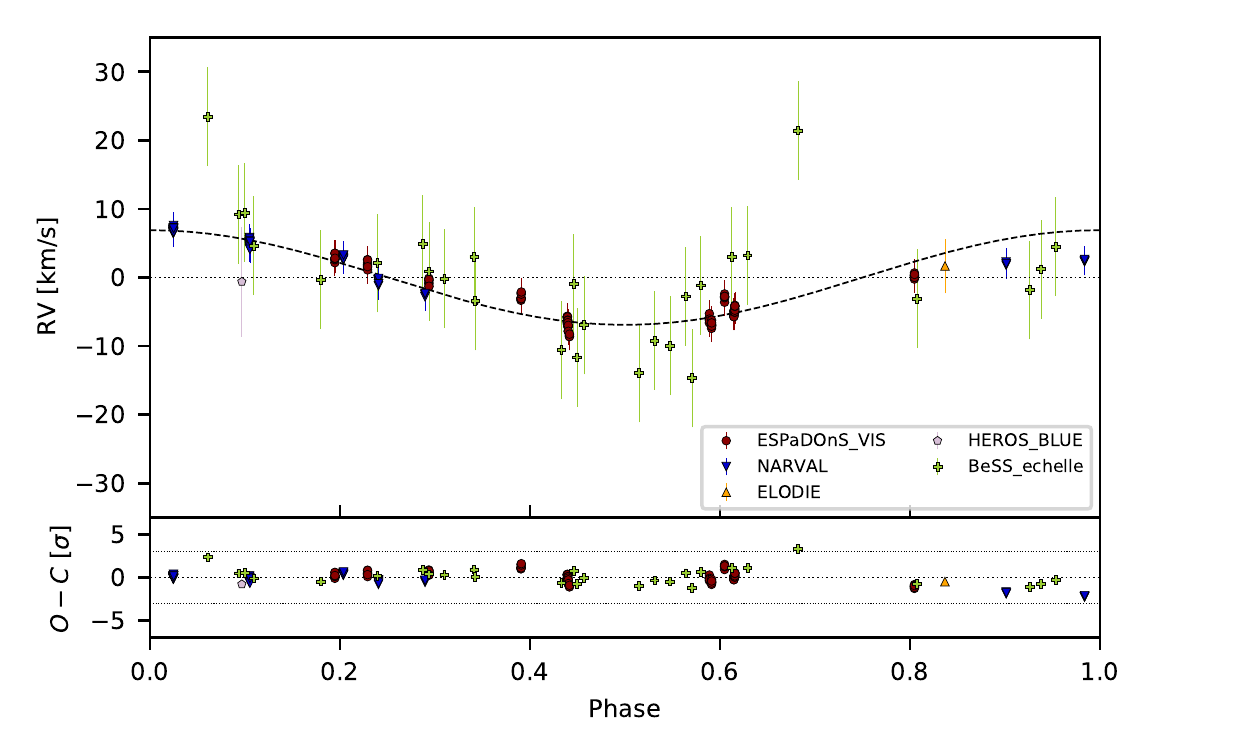}
   \caption{\label{fig:RVcurve_4026}RV curve of the primary component of $\kappa$~Dra with RV measurements from the \ion{He}{1}\,$\lambda4026$ line corrected for the zero point (different for each line) so that they are centered at 0\,km\,s$^{-1}$. The symbols identify different datasets as shown in the legend. The RV measurements for this line include the entirety of the ESPaDOnS and Narval datasets (the 168 spectra obtained in high cadence are clustered around individual observing nights), 28 BeSS echelle spectra, and a single spectrum each from {\sc HEROS} blue and ELODIE. The lower panel shows the $O-C$ residuals in units of $\sigma$. The full set of five figures with RV measurements from all five spectral lines is available in the online journal.
   }
\end{figure*}

In the next step, we included the full set of 1850 individual RV measurements. As expected, the combined RV curve shows considerable scatter for the lower-quality datasets. More specifically, the MUSICOS measurements in the \ion{He}{1}\,$\lambda4713$ line are systematically offset to positive values by about 15\,km\,s$^{-1}$.  To a smaller extent, the BeSS spectra (with mostly even phase coverage) show an offset to negative values for the \ion{He}{1}\,$\lambda4471$ line, while {\sc Heros} appears offset to positive values. However, since these systematics were limited to only these few instances and for lower-weighted measurements, they were not excluded from the solution. The resulting orbital parameters from the full dataset agree to within 1$\sigma$ of the first solution, while the uncertainty of $K_\mathrm{Be}$ slightly increased. Thus, the first solution using ESPaDOnS and Narval only is adopted as the final solution (Table~\ref{tab:orb_pars}). The final RV curve with RV measurements from \ion{He}{1}\,$\lambda4026$ is plotted in Fig.~\ref{fig:RVcurve_4026}. The full set of five figures with RV measurements from all five spectral lines is available in the online journal.

\begin{figure}
   \centering
   \includegraphics[width=0.5\textwidth]{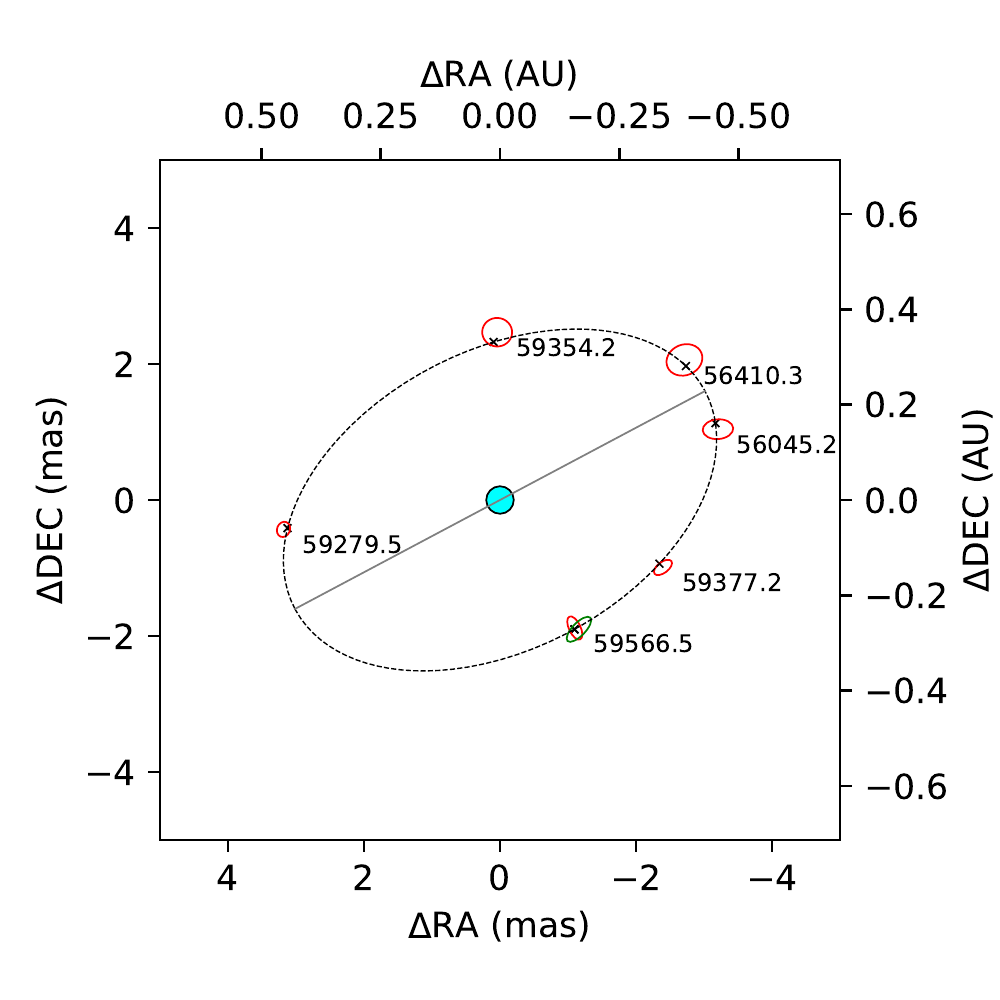}
   \caption{\label{fig:orbit}Relative astrometric orbit for the secondary component of $\kappa$~Dra. The measured positions are shown as error ellipses indicating $\pm5\sigma$ uncertainties (red for MIRC(-X) and green for MYSTIC) annotated with (HJD$-2400000.5$) dates. The corresponding points on the calculated orbit appear as crosses. The  UD of the central star (including any contribution from the disk) measured in 2021 in the $K$-band (0.405\,mas) is shown as a cyan filled circle.}
   \end{figure}

The astrometric orbit obtained is shown in Fig.~\ref{fig:orbit}. While there are only seven astrometric datapoints, the number of parameters constrained exclusively by the astrometry is only three - $i$, $a''$, and $\Omega$ - so that the size of the dataset is sufficient for the orbital solution. The median $(O-C)$ of the astrometric points is $76$\,$\mu$as, which is reasonable considering the high contrast resulting in the {\sc CP} signal close to zero, and the inability to fully use the other interferometric observables to constrain the companion position.

The orbital inclination is $130.0\pm3.4$\degree, indicating a projected motion from North through West, which is the clockwise direction in Fig.~\ref{fig:orbit}. The inclination of the circumstellar disk should then be close to 40{\degree} if the orbital and disk planes are aligned. The results from the disk fitting are consistent with this value, and the PA of the disk agrees well with $\Omega$ determined from the orbital solution (Table~\ref{tab:disk_params}). The resulting masses are discussed in Sect.~\ref{sec:nature_of_comp}.

\begin{deluxetable*}{lChCC}
\tablecaption{Final parameters compared to previous results\label{tab:orb_pars}}
\tablewidth{0pt}
\tablehead{
\nocolhead{} & \colhead{This work} & \nocolhead{all spectra} & \colhead{\cite{1991BAICz..42...39J}} & \colhead{\cite{2021RMxAA..57...91S}} 
}
\startdata
$P$ [d] & 61.5496 \pm 0.0058 & 61.5568 \pm 0.0038 & 61.5549\pm0.0032  & 61.55\pm0.02   \\
$T_{\rm RV max}$ [MJD] & 49980.4\pm1.3 & 49978.95\pm0.75 & 15757.52\pm0.99 & 49980.22\pm0.59   \\
$e$ & 0  & 0 & 0 & 0    \\
$\omega_\mathrm{Be}$ [\degree] & 0  & 0 & 0 & 0   \\
$K_\mathrm{Be}$ [km\,s$^{-1}$] & 6.90\pm0.15 & 6.81\pm0.21 & 8.29\pm0.48 & 6.81\pm0.24   \\
$q$ & 0.117\pm0.009 & 0.116\pm0.008 & 0.0744-0.1564 & \nodata   \\
$i$ [\degree] & 130.0\pm3.4 & 130.9\pm2.3 & \nodata & \sim30   \\
$\Omega$ [\degree] & 118.0\pm1.3 & 120.30\pm0.57 & \nodata & \nodata   \\
$a''$ [mas] & 3.414\pm0.001 & \nodata & \nodata   \\
$a$ [AU] & 0.487\pm0.021 & 0.489\pm0.021 & \nodata & \nodata   \\
$\gamma$ [km\,s$^{-1}$] & 12.0\pm0.3\tablenotemark{a} & 12.3\pm0.4\tablenotemark{b} & \nodata & \nodata \\
\hline
$M_\mathrm{Be}$ [\Msun] & 3.65\pm0.48 & 3.69\pm0.49 & 5.6 & 4.8\pm0.8   \\
$M_\mathrm{comp}$ [\Msun] & 0.426\pm0.043 & 0.430\pm0.041 & 0.0417-0.876 & \sim0.8 
\enddata
\tablenotetext{a}{Determined from the \ion{He}{1}\,$\lambda4713$ line.}
\end{deluxetable*}

\figsetstart
\figsetnum{3}
\figsettitle{RV curves.
}

\figsetgrpstart
\figsetgrpnum{3.1}
\figsetgrptitle{\ion{He}{1}\,$\lambda4026$}
\figsetplot{plot_RVcurve_kapDra_ESP+NARVAL_4026_5.0.pdf}
\figsetgrpnote{RV curve of the primary component of $\kappa$~Dra with RV measurements from the \ion{He}{1}\,$\lambda4026$ line corrected for the zero point (different for each line) so that they are centered at 0\,km\,s$^{-1}$. The symbols identify different datasets as shown in the legend. }
\figsetgrpend

\figsetgrpstart
\figsetgrpnum{3.2}
\figsetgrptitle{\ion{He}{1}\,$\lambda4471$}
\figsetplot{plot_RVcurve_kapDra_ESP+NARVAL_4471_5.0.pdf}
\figsetgrpnote{RV curve of the primary component of $\kappa$~Dra with RV measurements from the \ion{He}{1}\,$\lambda4471$ line corrected for the zero point (different for each line) so that they are centered at 0\,km\,s$^{-1}$. The symbols identify different datasets as shown in the legend. }
\figsetgrpend

\figsetgrpstart
\figsetgrpnum{3.3}
\figsetgrptitle{\ion{Mg}{2}\,$\lambda4481$}
\figsetplot{plot_RVcurve_kapDra_ESP+NARVAL_4481_5.0.pdf}
\figsetgrpnote{RV curve of the primary component of $\kappa$~Dra with RV measurements from the \ion{Mg}{2}\,$\lambda4481$ line corrected for the zero point (different for each line) so that they are centered at 0\,km\,s$^{-1}$. The symbols identify different datasets as shown in the legend. }
\figsetgrpend

\figsetgrpstart
\figsetgrpnum{3.4}
\figsetgrptitle{\ion{He}{1}\,$\lambda4713$}
\figsetplot{plot_RVcurve_kapDra_ESP+NARVAL_4713_5.0.pdf}
\figsetgrpnote{RV curve of the primary component of $\kappa$~Dra with RV measurements from the \ion{He}{1}\,$\lambda4713$ line corrected for the zero point (different for each line) so that they are centered at 0\,km\,s$^{-1}$. The symbols identify different datasets as shown in the legend. }
\figsetgrpend

\figsetgrpstart
\figsetgrpnum{3.5}
\figsetgrptitle{\ion{He}{1}\,$\lambda6678$}
\figsetplot{plot_RVcurve_kapDra_ESP+NARVAL_6678_5.0.pdf}
\figsetgrpnote{RV curve of the primary component of $\kappa$~Dra with RV measurements from the \ion{He}{1}\,$\lambda6678$ line corrected for the zero point (different for each line) so that they are centered at 0\,km\,s$^{-1}$. The symbols identify different datasets as shown in the legend. }
\figsetgrpend

\figsetend

\section{Fundamental parameters and nature of the companion}
\label{sec:nature_of_comp}

For the Be primary star, \citet{2005A&A...440..305F} derived $T_\mathrm{eff} = 13982\pm392$\,kK and $\log{g} = 3.479\pm0.061$ by fitting model atmospheres accounting for rapid rotation and gravitational darkening. \citet{2013ApJ...768..128T} fitted a Kurucz model spectrum \citep{1979ApJS...40....1K} with these parameters to UV spectra (IUE) to derive the limb-darkened disk (LD) diameter ($0.385\pm0.011$\,mas) and interstellar reddenning ($E(B-V) = 0.022\pm0.008$\,mag). The LD diameter was converted to a UD diameter of $0.381\pm0.011$\,mas using the linear limb-darkening parameter from \citet{2011A&A...529A..75C} and stellar parameters listed by \citet{2013ApJ...768..128T}. The UD diameter compares well with the interferometrically measured $H$-band UD diameter in an almost diskless state, which equals $0.354\pm0.004$ (Sect.~\ref{sec:cp_fitting}). However, since that measurement comes from a single calibrated interferometric bracket, the result is not very robust, so that in the following we use the former value determined from fitting the FUV spectrum. The average physical radius of $\kappa$~Dra implied by the Gaia DR3 distance is then $R_\mathrm{Be} = 5.85\pm0.18$\,\Rsun. From the above estimates of $R_\mathrm{Be}$ and $T_\mathrm{eff}$ for the primary Be star, the luminosity $L$ is $1178\pm151$\,\Lsun. The absolute $V$-band magnitude $M_V$ is $-1.95\pm0.40$\,mag resulting from the apparent magnitude $m_V$ of 3.89\,mag \citep{2002yCat.2237....0D}, an adopted $R_V$ of 3.1, and $E(B-V)$ determined by \citet{2013ApJ...768..128T}. This is in agreement with the typical $M_V$ for spectral types B6IIIe to B8IIIe, which are brighter by about 1.0--1.5\,mag than the corresponding non-Be stars of the same spectral types according to \citet{2006MNRAS.371..185W}. However, it should be noted that in that work the possible contribution from the disk was not accounted for, while it can amount to as much as $\sim0.5$\,mag \citep{2017AJ....153..252L}. The $v\sin{i}$ of $\kappa$~Dra is $200\pm12$\,km~s$^{-1}$ according to \citet{2005A&A...440..305F}.

The dynamical mass of the Be primary star obtained from the orbital parameters and Gaia DR3 distance is $3.65\pm0.48$\,\Msun\ (Sect.~\ref{sec:orb_solution}, Table~\ref{tab:orb_pars}). Comparing this value with the typical masses of slowly-rotating stars with similar $T_\mathrm{eff}$ and spectral type (B6) reveals that the mass of $\kappa$~Dra is slightly lower than expected. For instance, \citet{1988BAICz..39..329H} gives a mass range for B6 stars of $3.63$ to $3.93$\,\Msun\ with a mean of $3.78$\,\Msun, based on observed masses of 15 eclipsing binaries. Later compilations of masses of eclipsing binaries give still somewhat higher masses for stars with similar $T_\mathrm{eff}$, e.g., the secondary B6V component of U Oph has a mass of $4.58\pm0.05$\,\Msun\footnote{\url{https://www.astro.keele.ac.uk/jkt/debcat/}} \citep{2015ASPC..496..164S}. Relying on spectrophotometry and Hipparcos parallaxes, \citet{2010AN....331..349H} derived a median mass of $4.65\pm0.72$ for B6\,III stars. As for the radius, $R_\mathrm{Be} = 5.85\pm0.18$\,\Rsun\ corresponds to the equatorial radius, which is larger than the polar radius due to the rapid rotation. In the case of critical rotation, the polar radius is $R_\mathrm{p} = \frac{2}{3} R_\mathrm{Be} = 3.90\pm0.12$\,\Rsun, i.e., still significantly higher than the typical radius of B6 stars derived from eclipsing binaries, which is $\sim3.05$\,\Rsun\ \citep{1988BAICz..39..329H}.

As a crucial constraint for the nature of the companion, \citet{2017ApJ...843...60W} found from IUE spectra that the FUV photospheric flux ratio $f_\mathrm{phot, FUV}$ of a possible sdO companion has to be less than 1.0\% for the sdO star to remain undetected. Combining this information with the measured dynamical mass and the near-IR flux ratio then enables the characterization of the companion. To make use of the measured flux ratio, the contribution of the circumstellar disk (very small in 2021) should be taken into account. Estimating the exact values for the continuum contribution of the disk from the observed line emission is beyond the scope of this work, so that the average flux ratio measured in 2021 ($1.63\pm0.09$\% in the $H$~band) is taken as the photospheric flux ratio between the components $f_\mathrm{phot, H} = f_\mathrm{comp, phot, H} / f_\mathrm{Be, phot, H}$, neglecting the small contribution from the disk. Below, we consider the better-constrained $H$-band flux ratio, but the results are compatible when using the slightly lower $K$-band flux ratio.

\begin{figure*}
   \centering
   \includegraphics[width=1.0\textwidth]{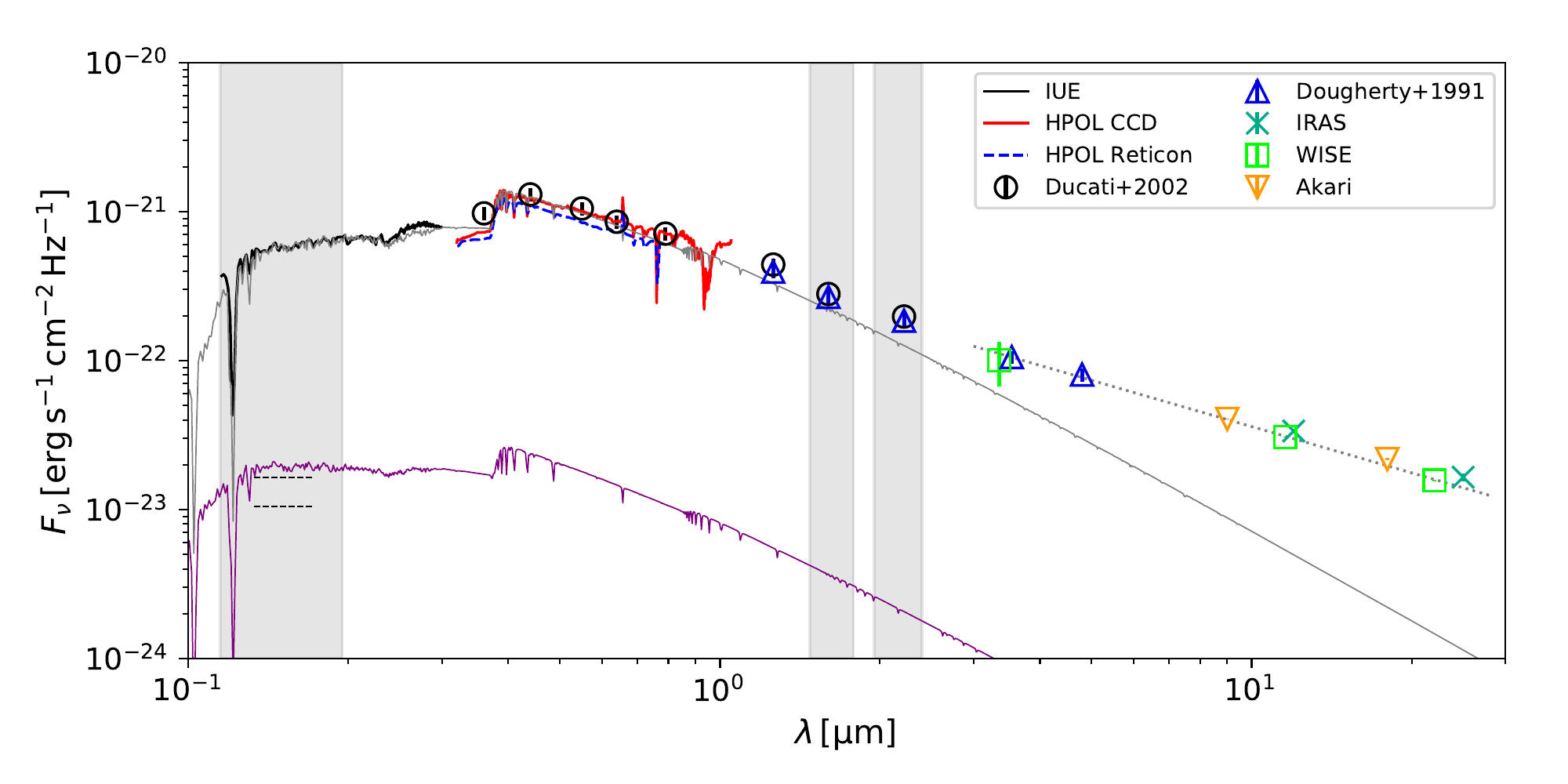}
   \caption{\label{fig:sed}Flux-calibrated spectrophotometry of $\kappa$~Dra overplotted in grey with a reddened ($E(B-V) = 0.022$ and $R_\mathrm{V} = 3.1$) Kurucz model spectrum representing the primary Be star ($T_\mathrm{eff} = 13982$\,kK, $\log{g} = 3.479$, $R = 5.85$\Rsun, $d=142.7$\,pc). The mid-IR power-law slope of the IR excess from the circumstellar disk (typical for Be stars) is shown as a dotted grey line. The reddened model spectrum for the sdB companion ($T_\mathrm{eff}=16.7$\,kK, $R=0.69$\,\Rsun) renormalized to the measured flux in the $H$~band (neglecting the possible contribution from the dissipating disk, see text) is shown in purple. The FUV flux ratio of $2.3\pm0.5$\% determined from FUV spectra is shown as two dashed black horizontal lines signifying the edges of the derived confidence interval. From short to long wavelengths, the grey vertical bands indicate the FUV region, the near-IR $H$~band, and $K$~band, respectively. }
\end{figure*}

In order to attempt to identify the nature of the companion, we used Kurucz/Castelli spectra\footnote{\url{https://www.stsci.edu/hst/instrumentation/reference-data-for-calibration-and-tools/astronomical-catalogs/castelli-and-kurucz-atlas}} \citep{2003IAUS..210P.A20C} to represent both the primary Be star ($T_\mathrm{eff} = 13982$\,kK, $\log{g} = 3.479$, and $R_\mathrm{Be} = 5.85$\,\Rsun) and the companion (Fig.~\ref{fig:sed}). The model spectra of the companion were renormalized to the measured companion flux in the $H$~band and chosen to represent either a stripped sdOB star or an MS star to obtain the expected flux ratio in the FUV $f_\mathrm{phot, FUV}$. 
This procedure also leads to an estimate of the radius of the companion, providing an additional constraint for the nature of the companion. This is done by comparing the surface fluxes from the Kurucz/Castelli models, which at the given $T_\mathrm{eff}$ would contribute the observed flux fraction in the $H$~band:
\begin{equation} \label{eq1}
    f_\mathrm{phot, H} = \frac{f_\mathrm{comp, phot, H}}{f_\mathrm{Be, phot, H}} = \frac{F_\mathrm{comp, phot, H}}{F_\mathrm{Be, phot, H}} \left( \frac{R_\mathrm{comp}}{R_\mathrm{Be}} \right)^2,
\end{equation}
where $F$ denotes fluxes per unit area, and $R_\mathrm{comp}$ is the companion's radius. This relation assumes that the flux from the companion does not suffer (variable) extinction from gas in the system.

First, values for $T_\mathrm{eff}$ of 25, 35, and 45\,kK, and for $\log{g}$ of 4.75 were used to represent an sdO companion. This results in an expected $f_\mathrm{phot, FUV}$ of $\sim$10, $\sim$20, and $\sim$25\%, respectively, which is clearly higher than the limit of 1\% \citep{2017ApJ...843...60W}. For a cooler sdB companion with $T_\mathrm{eff}$ of 15 and 20\,kK (and $\log{g}$ of 4.75), $f_\mathrm{phot, FUV}$ decreases to $\sim$2.3\% and $\sim$5.5\%, respectively. This implies that an sdO companion can be ruled out, while a cooler sdB remains a possibility, as the search in the FUV was optimized for an sdO rather than an sdB star \citep{2017ApJ...843...60W}. This means that the upper limit of $f_\mathrm{phot, FUV}$ of 1\% is probably too low in case of an sdB companion. 

A late-type MS star could be a possibility, as its contribution to the FUV flux would be negligible. Using the values given by \citet{2013ApJS..208....9P}\footnote{\url{http://www.pas.rochester.edu/~emamajek/EEM_dwarf_UBVIJHK_colors_Teff.txt}}, the measured dynamical mass of the companion best corresponds to an M2V star. However, the resulting $R_\mathrm{comp}$ for an M2V companion from Eq.~\ref{eq1} is $2.1$\,\Rsun, which is clearly too high for the expected radius of 0.446\,{\Rsun} \citep{2013ApJS..208....9P}. Thus, an MS companion appears to be ruled out as well.

We can also rule out the possibility of the companion being a WD due to the intrinsically extremely low luminosities of these stars of the order of $\sim10^{-2}$ to $\sim10^{-4}$\,{\Lsun} \citep[e.g.][]{2012ApJS..199...29G}. Even considering a bloated pre-WD, which would possibly have sufficient luminosity, would mean that it has to be much hotter than the primary, so that the same arguments described above for a hot sdO are valid. 

Given the evidence that the companion might be cooler than found for other stripped companions of Be stars, we re-inspected the ultraviolet spectral results from IUE that led to the null detection reported by \citet{2017ApJ...843...60W}. We followed the same method as before by forming cross-correlation functions (CCFs) in the wavelength range 1150--1950\,{\AA} of each spectrum with an assumed model template spectrum and then making a Doppler tomography reconstruction of the CCF components for the Be star and its companion.  However, the new analysis differs in three ways from that of \citet{2017ApJ...843...60W}. First, we omitted spectrum SWP29665, which leaves a set of 24 IUE high dispersion FUV spectra (see Sect.~\ref{sec:obs_sed}).  Second, we adopted the improved orbital elements and mass ratio estimate from Table~\ref{tab:orb_pars} to set the orbital velocities of the components used in the tomographic reconstruction. Finally, we used spectral model templates from the UVBLUE grid by \citet[][based upon the ATLAS9/SYNTHE codes by Robert Kurucz]{2005ApJ...626..411R} for $T_{\rm eff} = 13$ to 25~kK, $\log g = 5.0$, and solar metallicity. 

\begin{figure} 
   \centering
   \includegraphics[width=0.5\textwidth]{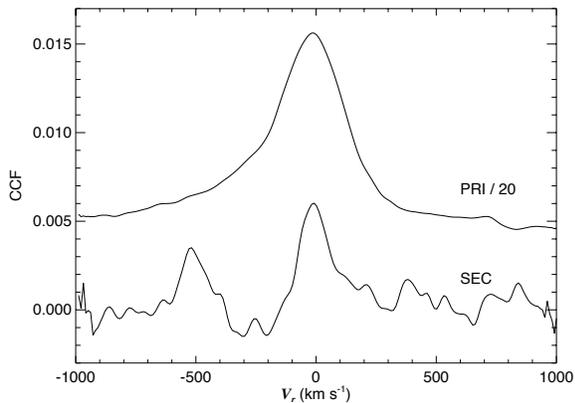}
   \caption{\label{fig:uvccf} The tomographic reconstructions of the CCFs for the Be star (PRI) and companion (SEC). The amplitude of the Be star CCF is reduced by a factor of 20 and offset by 0.005 for ease of comparison with the companion star CCF component. The signal from the companion is seen at a similar systemic velocity as the Be star (heliocentric frame), appears much narrower, and corresponds to a companion with a flux ratio of $2.3\pm0.5\%$ for the FUV range centered on 1500\,\AA .
   }
\end{figure}

This method using cooler model spectral templates reveals the spectral signature of the companion, so that the companion is spectroscopically detected for the first time, and the validity of our orbital solution (Table~\ref{tab:orb_pars}) is confirmed. This also marks the first time that a stripped sdB companion to a Be has been unambiguously detected. We show in Fig.~\ref{fig:uvccf} the resulting CCF components for the Be star and companion from the tomographic reconstruction using a model template with $T_{\rm eff} = 17$~kK. We detect a faint CCF peak for the companion that reaches a level about five times higher than the background standard deviation of the CCF (measured over the velocity range $600 < |V_r|< 1000$ km~s$^{-1}$). 
This peak appears narrow, suggesting a low $v\sin{i}$, as found for other stripped companions \citep{2021AJ....161..248W}, and it appears at a similar systemic velocity as measured for the Be star CCF component ($-11.0\pm1.1$ and $-14.8\pm1.2$\,km~s$^{-1}$ for the companion and Be star, respectively). The ratio of peak to background standard deviation for the companion CCF attains a maximum at $T_{\rm eff} = 16.7 \pm 2.0$~kK. We made a simulation of model binary spectra for the Be star ($T_{\rm eff} = 14$~kK, $\log = 3.5$, $V\sin i = 200$ km~s$^{-1}$) plus the companion for the same set of observed Doppler shifts and specific values of the monochromatic flux ratio $f_\mathrm{phot, FUV}$. We used the simulations to compare the ratio of CCF peaks in the models to those observed, and a match is obtained for a flux ratio of $f_\mathrm{phot, FUV} = 2.3\pm0.5\%$ at 1500\,\AA. This agrees with expectations based upon model atmospheres and the $H$-band stellar flux ratio obtained from interferometry (while neglecting a possible small contribution by the disk), which for the same $T_{\rm eff}$ predict an sdB radius of $R_\mathrm{comp} = 0.69\pm0.07$\,{\Rsun} and flux ratios of $f_\mathrm{phot, FUV} = 3.3\pm0.5\%$, and $f_\mathrm{phot, K} = 1.6\pm0.5\%$. Including a possible flux contribution by the almost-dissipated disk decreases the values of the predicted flux ratios, leading to a still improved agreement. The expected flux ratio in the visible ($\lambda=5000$\AA) is $\sim1.8$\%, so that it may be possible to detect the companion spectroscopically in the visible with renewed efforts.

\section{Discussion, conclusions, and outlook}
\label{sec:conclusions}

{\catcode`\&=11
\gdef\2005AandA...440..305F{\cite{2005A&A...440..305F}}}

\begin{deluxetable*}{cCCCCCC}
\tablecaption{Fundamental parameters of $\kappa$~Dra components}\label{tab:fundamental_params}
\tablewidth{0pt}
\tablehead{
\colhead{Component} & \colhead{$M$} & \colhead{$T_\mathrm{eff}$} & \colhead{$R$}  & \colhead{$L$} & \colhead{$v \sin{i}$} & \colhead{$M_V$}\\
\nocolhead{Component} & \colhead{[\Msun]} & \colhead{[K]} & \colhead{[\Rsun]} & \colhead{[\Lsun]} & \colhead{[km\,s$^{-1}$]} & \colhead{[mag]}
}
\startdata
Be & 3.65\pm0.48 & 13982\pm392\tablenotemark{a} & 5.85\pm0.18\tablenotemark{b}  & 1178\pm151 & 200\pm12\tablenotemark{a} & -1.95\pm0.40\tablenotemark{b}\\
sdB & 0.426\pm0.043 & 16700\pm2000 & 0.69\pm0.07 & 33\pm17 & 35\pm10 & 2.4\pm0.4\\
\enddata
\tablenotetext{a}{\2005AandA...440..305F}
\tablenotetext{b}{\cite{2013ApJ...768..128T} + Gaia DR3 parallax}
\end{deluxetable*}

Our state-of-the-art $H$- and $K$-band interferometry has directly detected the companion to the classical Be star $\kappa$~Dra. Together with the orbital RV amplitude of the Be primary measured in optical spectra and the parallax from Hipparcos and Gaia, these results enabled us to derive a full 3D orbital solution (Table~\ref{tab:orb_pars}), and the absolute masses and other fundamental parameters of both components (Table~\ref{tab:fundamental_params}). Using these results and the measured near-IR flux ratios between the components, we excluded a late-type MS star or a WD nature of the companion, while an evolved late-type star is ruled out by a lack of corresponding IR excess. In a targeted search, guided by the derived orbital solution and the previous non-detection in the FUV of an sdO companion, we successfully detected a stripped-down sdB companion, making $\kappa$~Dra the first firm case of a Be+sdB system, where the sdB companion has a substantially lower temperature than all the known sdO companions (found around hotter Be-star primaries).


Presently, the B6 primary star of $\kappa$~Dra is the lowest-mass Be star with a directly detected companion.  The sdB nature of the companion is in contrast with early-type Be binaries with detected stellar companions, among which sdO companions dominate (maybe even exclusively). This contrast may be due to a bias in the observational analyses of FUV spectra, in which the standard cross-correlation technique was optimized for the detection of sdO secondaries, which is the reason why the sdB companion to $\kappa$~Dra had not been previously detected \citep{2017ApJ...843...60W}.  The example of $\kappa$~Dra suggests that the search for hot low-mass companions in FUV spectra should be extended to the sdB domain, especially for cooler Be stars.  In combination with optical long-baseline interferometry, this can clarify whether or not there is any correlation between the mass of the Be primaries and the nature of those secondaries that are not NSs.  

\begin{figure} 
   \centering
   \includegraphics[width=0.5\textwidth]{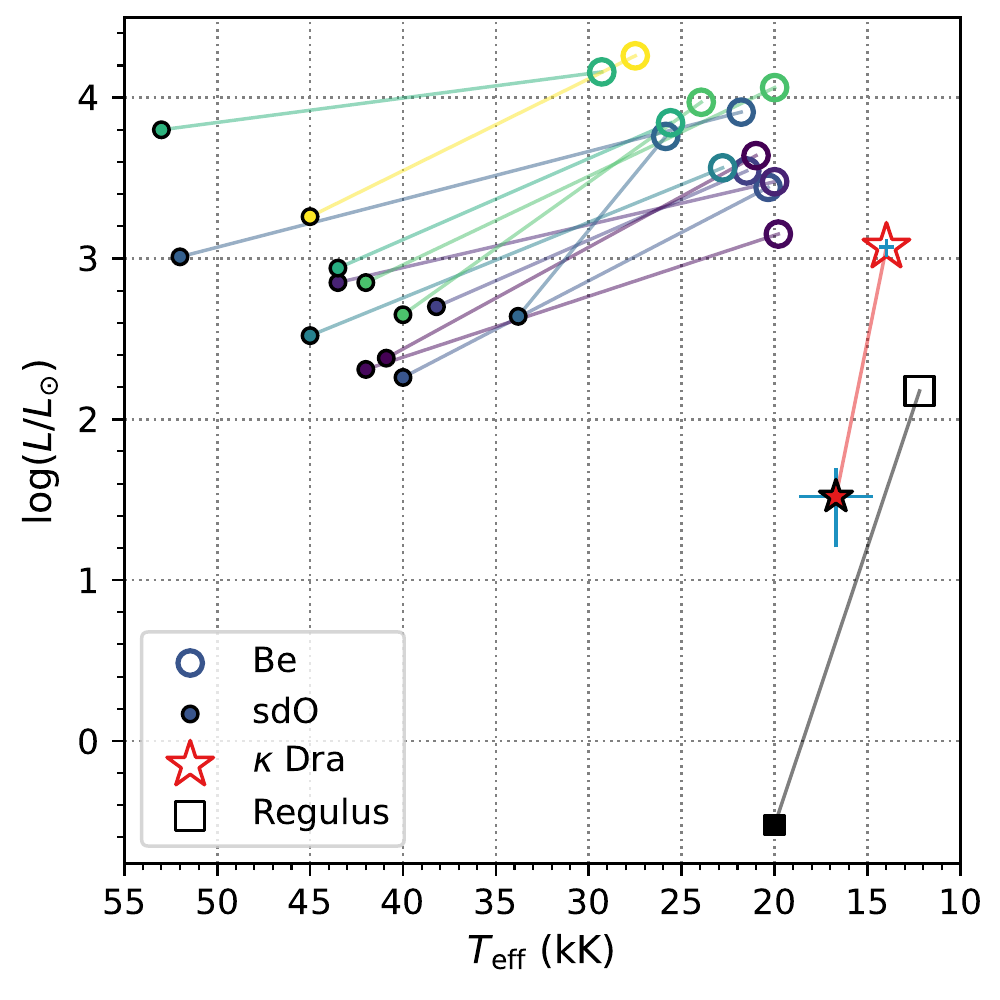}
   \caption{\label{fig:HRD} The luminosity and temperature of 13 known Be+sdO binaries are plotted as open circles (Be) and filled circles (sdO), with a line connecting the two components of each binary system. These parameters were estimated from this work and from \citet{2021AJ....161..248W}, \citet{2018ApJ...865...76C}, and \citet{2018A&A...615A..30S}. Each system is color-coded according to the mass of the Be component, with lighter colors indicating higher mass (maximum of 11.15\,\Msun, and minimum of 6.2\,\Msun). The Regulus system is plotted in a similar fashion with black squares. $\kappa$~Dra is plotted with star symbols, with errors as in Table~\ref{tab:fundamental_params}. 
   }
\end{figure}

A dependence of the type of the companion on the mass of the Be star is clear for NS companions which are not found around mid- or late-type Be stars (Sect.\,\ref{sec:intro}).  This is explained by the need for a massive progenitor of the secondary so that it could explode as a supernova. Fig.~\ref{fig:HRD} shows an HR diagram with the component parameters of the known (early) Be+sdO systems complemented by the B6e+sdB system $\kappa$~Dra and the B8n+pre-WD system Regulus. It appears plausible that there might be a hidden population of sdB or (pre-)WD companions around cooler Be stars, so that the nature of the companion indeed correlates with the mass of the primary Be star. The discovery of a stripped sdB companion should be a useful reference for binary stellar evolution models, as for instance the evolutionary grid of \citet{2018A&A...615A..78G} does not predict stripped companions with $T_\mathrm{eff} < 20$\,kK, and it expects a stripped star with a mass similar to the sdB companion of $\kappa$~Dra to have $T_\mathrm{eff} = 25.6$\,kK. However, it is possible that the companion to $\kappa$~Dra has already entered the WD cooling sequence.

The evidence that Be stars with subdwarf (and NS) companions owe their rapid rotation to mass and angular-momentum transfer in a close binary seems compelling.  However, for the characterization of the population of classical Be stars at the time of their formation, a larger number of physical parameter sets and also any non-detections of stellar companions of present-day Be stars are needed.  A critical bottleneck are the RV measurements.  For the RVs of the Be primaries, the large mass ratio and the strongly rotationally broadened photospheric lines require good data and careful analysis as the case of $\kappa$~Dra has confirmed.  RV determinations of subdwarfs orbiting Be stars are difficult \citep{2021AJ....161..248W} because of the large brightness difference.  The case of $\kappa$~Dra (in which statistical errors of $\sim$13\% and $\sim9$\% have been obtained for the masses of the primary and secondary, respectively) has illustrated how precise parallaxes can be used as a substitute for missing RVs of one of the two stars.  

For low-mass companions, the large flux ratio at the operating wavelengths of Hipparcos and Gaia becomes an advantage because the effect of the orbital motions on the astrometric solution is less detrimental.  But very high fluxes saturate the detectors and, contrary to first intuition, do not automatically promise larger parallaxes.  According to the BeSS database, the two dozen Be stars that are visually brighter than $\kappa$~Dra are mostly of earlier spectral type, and their higher intrinsic brightness implies distances that can actually be larger.  This double disadvantage is alleviated for stars similar to $\kappa$~Dra in both spectral type and apparent magnitude, of which well over a dozen exist.  Studies of these lower-mass Be stars can complement the more-massive-Be-star population with detected sdO stars.  A more representative sample of Be stars and subdwarf companions will deliver important input to models for the evolution of close binaries.  Any non-detections may have a bearing on the fraction of Be stars, if any, that do not form through mass transfer in a binary.  

$\kappa$~Dra is only the fifth Be binary for which dynamical masses have been determined (Sect.\,\ref{sec:intro}).  Of these, the Be stars in $\phi$~Per \citep{2015A&A...577A..51M} and V2119~Cyg \citep{2022ApJ...926..213K} have masses that are in agreement with their spectral types.  The Be star in the triple system $\nu$~Gem (for this star, a compact companion is not known) is of the same spectral type as $\kappa$~Dra, but at $3.33\pm0.10$~\Msun\ \citep{2021ApJ...916...24K} it is even slightly more undermassive than $\kappa$~Dra.  Something similar probably holds for 60~Cyg, in which the Be star has a dynamical mass of $7.7\pm1.2$~\Msun\ \citep[Klement, unpublished update of ][based on new observations]{2022ApJ...926..213K} whereas its spectral type B1\,Ve \citep{1982ApJS...50...55S} suggests 13.5\,\Msun\ and $11.98\pm1.7$\,\Msun\ for the calibrations of \citet{1988BAICz..39..329H} and \citet{2010AN....331..349H}, respectively.   At the current low level of knowledge and understanding, undermassive appears crudely equivalent to overluminous.  This is is not implausible for mass gainers which also gained thermal energy through the accretion and may experience enhanced mixing if the added gas has a higher molecular weight.  Furthermore, more advanced evolution may contribute to the overluminosity.  To assess the possibility of systematic differences between Be and non-Be stars in their fundamental parameters will be an important objective of future interferometry of Be stars.  As this study of $\kappa$~Dra has shown, the availability of radial velocities with matching quality can be decisive.

Of high interest are also the parameters that may trace the evolution of $\kappa$~Dra as a system, beyond that of its individual components.  The vanishingly small eccentricity is in agreement with orbital circularization due to tidal forces during the mass transfer. The disk appears to be coplanar with the orbit as would be expected, but high-resolution spectro-interferometry across prominent emission lines would be needed to deduce whether the angular-momentum vectors of the disk and the orbit are parallel or antiparallel.  In the latter case, one might wonder whether an initial third body, for which there is no evidence, could link such a peculiarity and the relatively high space velocity (see Sect.\,\ref{sec:observations_hip_gaia}).

\begin{acknowledgments}
This work is based upon observations obtained with the Georgia State University Center for High Angular Resolution Astronomy Array at Mount Wilson Observatory. The CHARA Array is supported by the National Science Foundation under Grant No.\ AST-1636624, and AST-2034336. Institutional support has been provided from the GSU College of Arts and Sciences and the GSU Office of the Vice President for Research and Economic Development.
MIRC-X received funding from the European Research Council (ERC) under the European Union's Horizon 2020 research and innovation programme (Grant No.\ 639889) and N.A., C.L.D., and S.K.\ acknowledge funding from the same grant.
Based on INES data from the IUE satellite.
This research has made use of the SIMBAD database, operated at CDS, Strasbourg, France.
This work has made use of the BeSS database, operated at LESIA, Observatoire de Meudon, France: http://basebe.obspm.fr.
This research has made use of NASA’s Astrophysics Data System.
This work has made use of data from the European Space Agency (ESA) mission
{\it Gaia} (\url{https://www.cosmos.esa.int/gaia}), processed by the {\it Gaia}
Data Processing and Analysis Consortium (DPAC,
\url{https://www.cosmos.esa.int/web/gaia/dpac/consortium}). Funding for the DPAC
has been provided by national institutions, in particular the institutions
participating in the {\it Gaia} Multilateral Agreement.
R.K.\ is grateful for a postdoctoral associateship funded by the Provost’s Office of Georgia State University. The research of R.K.\ is also supported by the National Science Foundation under Grant No.\ AST-1908026.
This research has made use of the Jean-Marie Mariotti Center \texttt{Aspro}, \texttt{OIFits Explorer}, and \texttt{SearchCal} services (available at http://www.jmmc.fr/aspro and http://www.jmmc.fr/searchcal).
R.K.\ expresses sincere thanks to the CHARA science staff, namely Matthew D.\ Anderson, Theo ten Brummelaar, Christopher Farrington, Rainer Koehler, Robert Ligon, Olli Majoinen, Nicholas Scott, Judit Sturmann, Laszlo Sturmann, Nils Turner, and Norman Vargas for making the observations used in this study.
A.C.C. acknowledges support from CNPq (grant 311446/2019-1) and FAPESP (grants 2018/04055-8 and 2019/13354-1). 
S.K.\ acknowledges support from an ERC Consolidator Grant (Grant Agreement ID 101003096) and STFC Consolidated Grant (ST/V000721/1).
J.D.M.\ acknowledges funding for the development of MIRC-X (NASA-XRP NNX16AD43G, NSF-AST 1909165) and MYSTIC (NSF-ATI 1506540, NSF-AST 1909165).
P.T.\ acknowledges CAPES funding, under the grant 88887.604774/2021-00.
\end{acknowledgments}

\facilities{Akari, CFHT, CHARA, Gaia, HIPPARCOS, IRAS, IUE, OHP:1.93m, OO:2, TBL, WISE}

\bibliography{kapDra}{}
\bibliographystyle{aasjournal}

\end{document}